\documentclass[draft]{agujournal2019}
\usepackage{url} %this package should fix any errors with URLs in refs.
\usepackage{lineno}
\usepackage{amsmath}
\usepackage{soul}
\usepackage[inline]{trackchanges} %for better track changes. finalnew option will compile document with changes incorporated.
\draftfalse

\journalname{Journal of Advances in Modeling Earth Systems (JAMES)}

\begin{document}

%% ------------------------------------------------------------------------ %%
%  Title
%
% (A title should be specific, informative, and brief. Use
% abbreviations only if they are defined in the abstract. Titles that
% start with general keywords then specific terms are optimized in
% searches)
%
%% ------------------------------------------------------------------------ %%

%% ------------------------------------------------------------------------ %%
%
%  AUTHORS AND AFFILIATIONS
%
%% ------------------------------------------------------------------------ %%

% Authors are individuals who have significantly contributed to the
% research and preparation of the article. Group authors are allowed, if
% each author in the group is separately identified in an appendix.)

% List authors by first name or initial followed by last name and
% separated by commas. Use \affil{} to number affiliations, and
% \thanks{} for author notes.
% Additional author notes should be indicated with \thanks{} (for
% example, for current addresses).

%\title{Bridging Intelligibility and Accuracy: A Simple Spectral Radiation Scheme for Climate Models}

%\title{A Simple Spectral Radiation Scheme for Idealized Climate Modeling}

\title{Bridging Clarity and Accuracy: A Simple Spectral Longwave Radiation Scheme for Idealized Climate Modeling}

%\title{A Simple Spectral Longwave Radiation Scheme for Idealized Climate Modeling}

%\title{Bridging Accuracy and Intelligibility: A Simple Spectral Radiation Scheme for Idealized Climate Modeling}

\authors{Andrew I.L. Williams$^{1}$}

\affiliation{1}{Program in Atmospheric and Oceanic Sciences, Princeton University, Princeton, NJ}

% \affiliation{1}{First Affiliation}
% \affiliation{2}{Second Affiliation}
% \affiliation{3}{Third Affiliation}
% \affiliation{4}{Fourth Affiliation}
%\affiliation{1}{}

%% Corresponding Author:
% Corresponding author mailing address and e-mail address:

% (include name and email addresses of the corresponding author.  More
% than one corresponding author is allowed in this LaTeX file and for
% publication; but only one corresponding author is allowed in our
% editorial system.)

% Example: \correspondingauthor{First and Last Name}{email@address.edu}

\correspondingauthor{Andrew Williams}{andrew.williams@princeton.edu}

%% Keypoints, final entry on title page.

%  List up to three key points (at least one is required)
%  Key Points summarize the main points and conclusions of the article
%  Each must be 100 characters or less with no special characters or punctuation and must be complete sentences

% Example:
% \begin{keypoints}
% \item	List up to three key points (at least one is required)
% \item	Key Points summarize the main points and conclusions of the article
% \item	Each must be 100 characters or less with no special characters or punctuation and must be complete sentences
% \end{keypoints}

\begin{keypoints}
\item We introduce an intermediate-complexity longwave radiative transfer scheme for climate models, the Simple Spectral Model (SSM).

\item The SSM captures the spectral dependence of atmospheric absorption using analytic fits, and is easy to understand and modify.

\item The SSM produces accurate climate states, and their response to warming, when implemented in a GCM, in stark contrast to gray schemes.

%Other RT codes can balance accuracy and efficiency, but intelligibility / transparency is often missing. 
%Makes it hard to understand the model that you're using, and to edit it in a way which makes sense for your application.
%Many studies have used gray radiation schemes, which are simple and analytically tractable, but suffer from large biases in radiative heating rates and other climate features. 
%Recent years have seen many advances in our understanding of radiative transfer using `simple spectral models' (SSMs). These have changed how we understand and teach many aspects of radiative transfer, but to date there has been no attempt to use this simplified approach to actually \textit{simulate} climate and climate change. Our goal here will be to fill this gap, and construct a SSM for GCMs
\end{keypoints}

%% ------------------------------------------------------------------------ %%
%
%  ABSTRACT and PLAIN LANGUAGE SUMMARY
%
% A good Abstract will begin with a short description of the problem
% being addressed, briefly describe the new data or analyses, then
% briefly states the main conclusion(s) and how they are supported and
% uncertainties.

% The Plain Language Summary should be written for a broad audience,
% including journalists and the science-interested public, that will not have 
% a background in your field.
%
% A Plain Language Summary is required in GRL, JGR: Planets, JGR: Biogeosciences,
% JGR: Oceans, G-Cubed, Reviews of Geophysics, and JAMES.
% see http://sharingscience.agu.org/creating-plain-language-summary/)
%
%% ------------------------------------------------------------------------ %%

%% \begin{abstract} starts the second page

\begin{abstract}
Parameterizing radiative transfer in means navigating trade-offs between physical accuracy and conceptual clarity. However, currently available schemes sit at the extremes of this spectrum: correlated-k schemes are fast and accurate but rely on lookup tables which obscure the underlying physics and make such schemes difficult to modify, while gray radiation schemes are conceptually straightforward but introduce significant biases in atmospheric circulation. Here we introduce a Simple Spectral Model (SSM) for clear-sky longwave radiative transfer which bridges this `clarity-accuracy' gap. The SSM accomplishes this by representing the spectral structure of H$_{2}$O and CO$_{2}$ absorption using analytic fits at reference conditions, then uses simple functional forms to extend these fits to different atmospheric conditions. This, coupled to a simple, two-stream solver, yields a system of six equations and ten physically-meaningful parameters which can solve for clear-sky longwave fluxes given atmospheric profiles of temperature and humidity. When implemented in an idealized aquaplanet GCM, the SSM produces zonal-mean climate states which accurately mimic the results using a benchmark correlated-k code. The SSM also alleviates the significant zonal-mean climate biases associated with using gray radiation, including an improved representation of radiative cooling profiles, tropopause structure, jet dynamics, and Hadley Cell characteristics both in control climates and in response to uniform warming. This work demonstrates that even a simple spectral representation of atmospheric absorption suffices to capture the essential physics of longwave radiative transfer, and the SSM promises to be a valuable tool both for idealized climate modeling, and for teaching radiative transfer in the classroom. 
\end{abstract}

\section*{Plain Language Summary}
The Earth is warmed by the sun, and cooled by emitting light out to space. This emission from the planet's surface is complicated by the existence of greenhouse gases, which absorb and re-emit the Earth's light. Representing these processes in climate models is important to successfully model the distribution of temperature and winds. Through decades of work we can now represent these processes to a high degree of accuracy, but the way we do this is difficult to understand unless you are an expert. Other methods exist which are easy to understand, but they do not give good representations of temperature or winds. In this paper I introduce a new method for representing the absorption and emission of light in the atmosphere. It is only a couple of equations and is easy to understand and to code up, but it is also very accurate and gives good representation of temperature and winds. This new approach is not meant to replace existing state-of-the-art methods, but rather to `fill in the gap' which exists between accurate methods which are hard-to-understand, and easy-to-understand methods which are inaccurate.

%% ------------------------------------------------------------------------ %%
%
%  TEXT
%
%% ------------------------------------------------------------------------ %%

%%% Suggested section heads:
 \section{Introduction}

Radiative transfer lies at the heart of Earth's climate system, governing the fundamental energy balance that drives atmospheric circulation and the hydrological cycle. Parameterizing radiation accurately and efficiently is therefore essential for developing useful general circulation models (GCMs). Line-by-line models exist which agree excellently with observations \cite{pincus2015radiative} but these benchmark models are much too computationally expensive to be used in GCMs. This creates an inevitable `accuracy-efficiency' trade-off that must be navigated in climate model development.

Beyond computational considerations, another critical factor must be balanced when developing radiative transfer schemes: their clarity and intelligibility. This axis of consideration follows from Isaac Held's influential 2005 essay where he cautioned against the growing `gap between simulation and understanding' in climate modeling \cite{held05b}. Even if parameterization schemes are accurate, a lack of clarity hampers our ability to design creative experiments and test hypotheses \cite{jeevanjee17, reed2025idealized}. It is this trade-off between clarity and accuracy that we will focus on in this paper, with an eye towards developing a simple radiative transfer scheme which balances these two constraints.

\begin{figure}
\centering
 \noindent\includegraphics[width=0.6\textwidth]{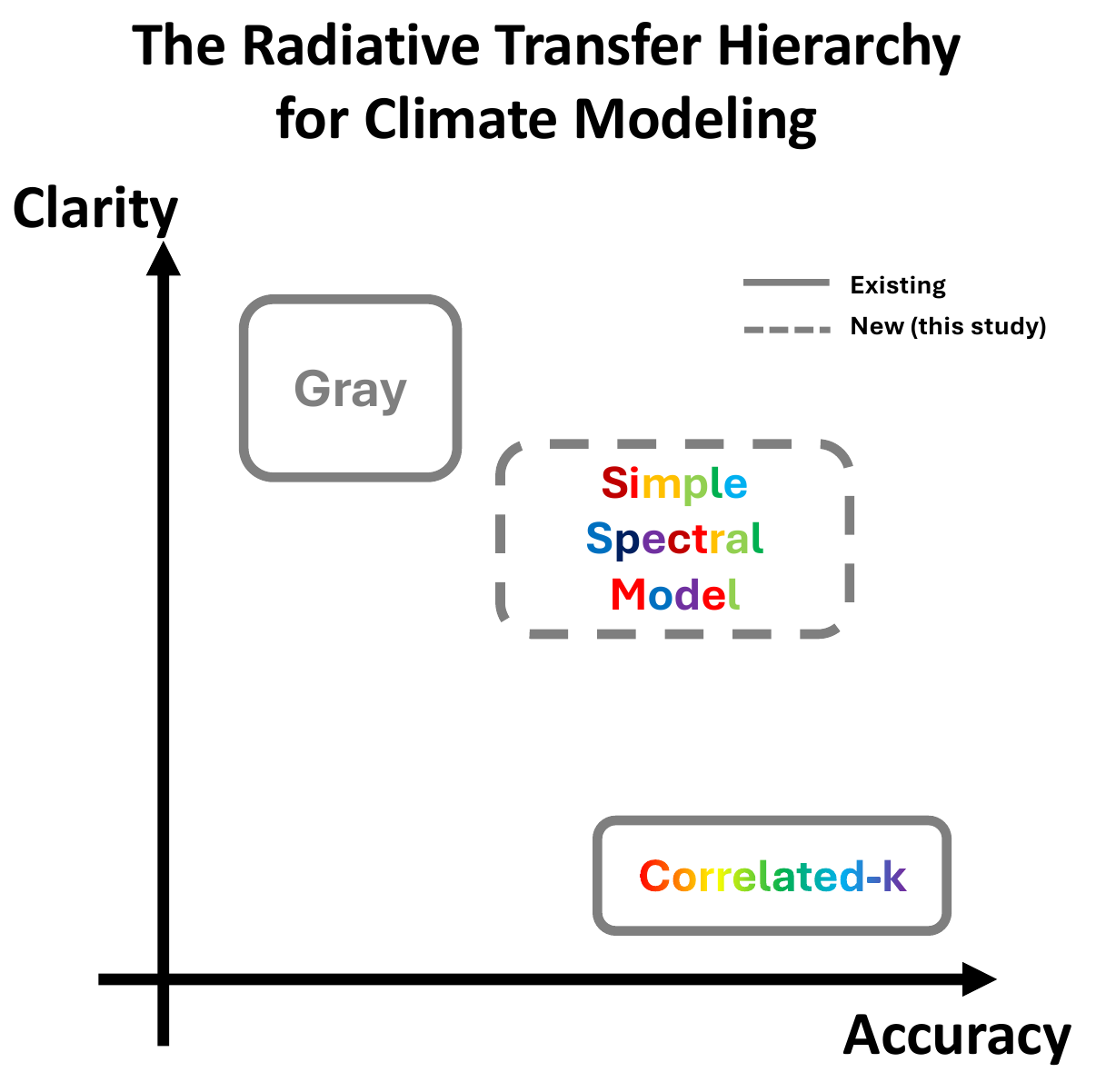}
\caption{A planar hierarchy of radiative transfer approaches for climate modeling. The different classes of radiative transfer schemes are grouped by their `clarity' (vertical axis) and their `accuracy' (horizontal axis). Both terms are used loosely, in a way which is defined in the main text. The Simple Spectral Model and correlated-k schemes are written using a rainbow font to highlight that they are \textit{spectral} representations of radiative transfer, as opposed to the gray schemes, which do not explicitly represent the spectral nature of light.}
\label{fig-introschematic}
\end{figure}

The current state of affairs, focusing on the representation of gas absorption, is sketched in Figure \ref{fig-introschematic}, which groups different classes of radiative transfer schemes according to their `clarity' and `accuracy'. Modern climate models typically parameterize gas absorption using correlated-k methods, such as RRTMG \cite{mlawer97, iacono2008radiative} and its successor RTE+RRTMGP \cite{pincus2019balancing}. These schemes typically split the electromagnetic spectrum into bands and use precomputed lookup tables to represent the spectral variation of gaseous absorption in a computationally efficient way. Correlated-k schemes can achieve high accuracy when applied within their calibration range \cite{pincus2015radiative}, and thus score highly on the `accuracy' axis. 

However, this accuracy often comes at the cost of clarity. The extensive use of lookup tables can obscure the underlying physics and means that modifying these schemes typically requires specialist knowledge, limiting their accessibility for hypothesis-driven experimentation or educational purposes \cite{hogan2020evaluating}. Furthermore, correlated-k schemes often provide only broadband output or spectral output aggregated into coarse bands, limiting the ability to link model output to spectrally resolved radiative processes. Hence, while correlated-k schemes represent the state-of-the-art for radiative transfer in GCMs, their complexity can pose challenges for interpretability and pedagogical use.

%As Isaac Held noted in his influential 2005 essay, climate modeling faces `a growing gap between high-end simulations and idealized theoretical work,' where increasingly sophisticated models risk becoming opaque worlds unto themselves, disconnected from physical intuition. 

Sitting at the opposite end of the `clarity-accuracy' spectrum we have `gray' radiation schemes. Gray radiation schemes neglect the spectral dependence of atmospheric absorption and emission, and in practice often directly prescribe analytic profiles of optical depth as a function of pressure. This greatly simplifies the mathematics of radiative transfer, offering conceptual clarity, analytical tractability, and the ability to easily modify the scheme to test hypotheses \cite<e.g.,>{lutsko2018influence}. This high score on the `clarity' axis (Fig. \ref{fig-introschematic}) has spurred the wide adoption of gray radiation schemes (most notably the \citeA{Frierson06a} scheme) in idealized GCM studies, yielding insights into the hydrological cycle \cite{Kang08,OGorman08b,Schneider10}, atmospheric dynamics \cite{wills17,davis2022eddy, lewis2024assessing}, and exoplanets \cite{merlis10, kaspi15, guendelman20}. However, it is well-known that gray schemes do not accurately represent the structure of radiative heating in the atmosphere \cite<e.g.,>{jeevanjee2020simple} and there is increasing realization in the community that gray radiation distorts many aspects of the atmospheric circulation and its response to warming \cite{tan19, davis2022eddy}. Hence, although gray schemes are conceptually simple and easy-to-use, they score poorly on the `accuracy' scale in Figure \ref{fig-introschematic}.

%This high score on the `clarity' axis (Fig. \ref{fig-introschematic}) has spurred the wide adoption of gray radiation schemes (most notably the \citeA{Frierson06a} scheme) in idealized GCM studies, yielding insights into the hydrological cycle \cite{Kang08,OGorman08b,Schneider10}, monsoons \cite{wei2018energetic, geen2019processes, lutsko2019modulation}, atmospheric dynamics \cite{ogorman08c,merlis11,levine15,wills17,davis2022eddy, lewis2024assessing}, the coupled climate system \cite{tuckman2025understanding,tuckman2025enso}, and exoplanets \cite{merlis10, schneider2009formation, kaspi15, guendelman20, Showman13}. 

As Figure \ref{fig-introschematic} illustrates, there is a conspicuous gap in the middle ground—we lack a radiation scheme that combines reasonable accuracy with reasonable clarity. Recent advances in `simple spectral models' (SSMs) offer an opportunity to fill this gap. Work by \citeA{koll18, jeevanjee2020simple, romps2022forcing, cohen2025spectroscopic}, among others, has shown that simplified representations of H$_{2}$O and CO$_{2}$ spectroscopy can capture the essential physics while remaining analytically tractable. This `pencil-and-paper' approach has yielded valuable insights into radiative forcing and feedbacks, hydrological sensitivity, and the vertical structure of radiative cooling. However, despite their theoretical utility, simplified spectroscopic models of radiative transfer have not yet been implemented in GCMs.

Here, we address the `clarity-accuracy' gap in Figure \ref{fig-introschematic} by developing a simple spectral model (SSM) for clear-sky, longwave radiative transfer in a GCM. The SSM combines the two-stream equations with analytical approximations to absorption coefficients, providing the spectral detail required for accurate radiative transfer while retaining analytical simplicity. When implemented in an idealized aquaplanet GCM, the SSM greatly reduces biases present when using a representative gray scheme \cite{Frierson06a}.

We focus on the longwave as it is the dominant source of radiative heating in Earth-like atmospheres, and as previous work on SSMs has focused mostly on the longwave. However, we view the current work only as a first step towards filling the `clarity-accuracy' gap in Figure \ref{fig-introschematic}, and as a complement to more advanced schemes like RTE+RRTMGP that prioritize accuracy and flexibility through different means.

We proceed as follows: In the next section, we will document the Simple Spectral Model (SSM) in terms of its governing equations and analytic approximations. Then we implement a hierarchy of longwave radiative transfer schemes (gray, SSM, RRTMG) in an idealized aquaplanet GCM and describe the resulting control climates and their response to uniform warming. We then use a single-column model to explore state-dependence of the longwave radiative feedback parameter in the SSM, with a focus on extremely warm climate states. The conclusion discusses paths forward for and applications of the SSM.

\section{Constructing the Simple Spectral Model}

\subsection{Overview}

In this section we will construct a simple spectral model (SSM) for clear-sky, longwave radiative transfer. Our SSM computes upward and downward longwave fluxes using the standard Schwartzchild's equations for flux in the presence of emission and absorption:

\begin{equation}
    \frac{dF_{\tilde{\nu}}^{\uparrow}}{d\tau_{\tilde{\nu}}} = F_{\tilde{\nu}}^{\uparrow} - \pi B_{\tilde{\nu}}(T)
    \label{eqn-twostreamLWup}
\end{equation}

\begin{equation}
    \frac{dF_{\tilde{\nu}}^{\downarrow}}{d\tau_{\tilde{\nu}}} = \pi B_{\tilde{\nu}}(T) - F_{\tilde{\nu}}^{\downarrow} 
    \label{eqn-twostreamLWdn}
\end{equation}

\noindent where $T$ is the temperature, and $\tau_{\tilde{\nu}}$ and $B_{\tilde{\nu}}(T)$ are the longwave optical depth and Planck function at wavenumber $\tilde{\nu}$, respectively. The boundary condition at the top of the atmosphere is $F_{\tilde{\nu}, \mathrm{ TOA}}^{\downarrow}=0$, and the boundary condition at the surface is $F_{\tilde{\nu}, \mathrm{ sfc}}^{\uparrow}=\pi B_{\tilde{\nu}}(T_{\mathrm{sfc}})$. We assume a unit surface emissivity, but this could easily be adjusted.

In addition to the boundary conditions, to solve Eqns. (\ref{eqn-twostreamLWup}) and (\ref{eqn-twostreamLWdn}) we also need to know the optical depth, $\tau_{\tilde{\nu}}(p)$. We calculate this as

\begin{equation}
    \tau_{\tilde{\nu}}(p)=D\int_{0}^{p}\kappa(\tilde{\nu}, T, p) \, q_{\mathrm{GHG}} \frac{dp'}{g}
    \label{eqn-opticaldepth}
\end{equation}

\noindent where $D=1.5$ is a diffusivity factor required by the two-stream approximation \cite{armstrong1968theory, clough1992line,Pierrehumbert10}. $\kappa$ is the mass absorption coefficient, which depends not only on wavenumber $\tilde{\nu}$ but also on $T$ and $p$, due to temperature scaling and pressure broadening. $q_{\mathrm{GHG}}$ is the mass-specific concentration (units of kg kg$^{-1}$) of the longwave absorber, which for H$_{2}$O is the specific humidity, $q_{v}$. For well-mixed CO$_{2}$, $q_{\mathrm{CO}_{2}}$ is the ppmv of CO$_{2}$ multiplied by the ratio of the molar masses of CO$_{2}$ and dry air; e.g. for 280 ppmv of CO$_{2}$, $q_{\mathrm{CO}_{2}}=280\times 10^{-6} \times (44/29)$.

%As our focus in this paper is on a simplified representation of longwave radiative transfer, we assume that the atmosphere is transparent in the shortwave, such that the shortwave radiative heating rate is zero in the atmosphere. The TOA insolation is prescribed as a function of latitude, $\phi$, by

%\begin{equation}
%    S_{\mathrm{TOA}}(\phi) = \frac{S_{0}}{4}\left[1+\frac{\Delta_{s}}{2}\left(1-3\sin^{2}\left(\phi\right)\right)\right]
%    \label{eqn-insol}
%\end{equation}

%\noindent which represents perpetual equinox conditions, with no diurnal cycle. This flavor of idealized shortwave radiative transfer is common in idealized GCM studies (citations), and so we adopt it here. However, it is possible to also develop simple spectral models of shortwave radiative transfer, and we will return to this point in the Discussion section. 

\subsection{Analytically approximating the absorption coefficients}

\begin{figure}
 \noindent\includegraphics[width=\textwidth]{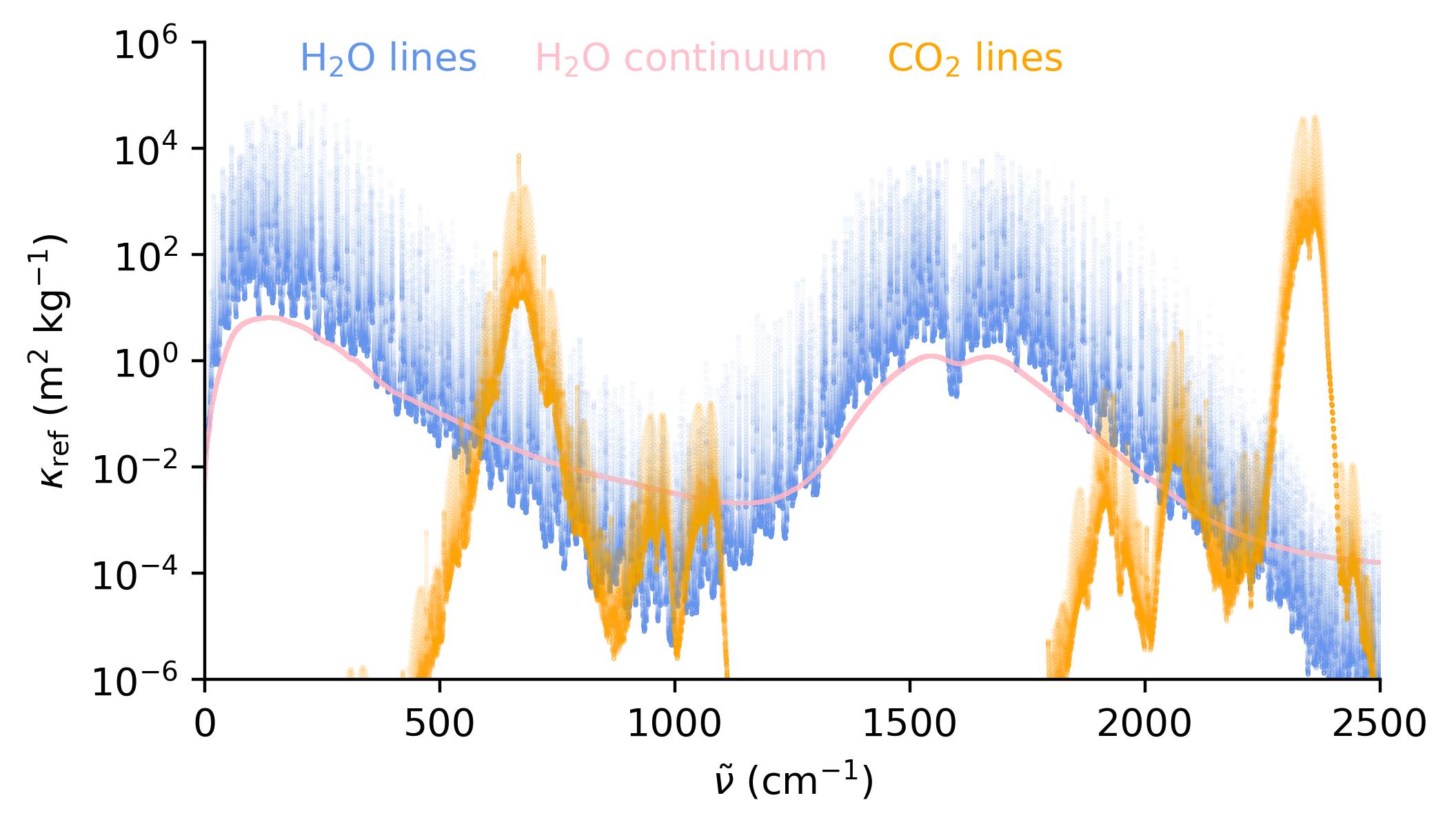}
\caption{Reference mass absorption coefficients (m$^{2}$ kg$^{-1}$) for H$_{2}$O line absorption (light blue), continuum absorption (light pink) and CO$_{2}$ line absorption (orange) at $\left(T_{\mathrm{ref}}, \,  p_{\mathrm{ref}}\right)$ = (260K, 500hPa) and 100\% relative humidity calculated using PyRADS \cite{koll18}.}
\label{fig-kappaTrue}
\end{figure}

To calculate the spectrally-resolved optical depth, $\tau_{\tilde{\nu}}$ we need to know the mass absorption coefficient, $\kappa_{\tilde{\nu}}$, but what does this look like? In Figure \ref{fig-kappaTrue}, we plot the mass absorption coefficients for H$_{2}$O line absorption (in blue), continuum absorption (in pink) and CO$_{2}$ line absorption (in orange) as a function of wavenumber. Line data are taken from the HITRAN2016 spectroscopic database \cite{gordon2017hitran2016} and the H$_{2}$O continuum is represented using the MT\_CKD\_3.2 model \cite{mlawer2012development}. The absorption coefficients are calculated at a reference pressure of 500hPa, a temperature of 260K, and 100\% relative humidity using the PyRADS model \cite{koll18}. The data are sampled at a spectral resolution of 0.01cm$^{-1}$, and line data plotted as partially-transparent dots such that darker colors correspond to a higher density of absorbing wavenumbers.

The complexity of the spectral features in Figure \ref{fig-kappaTrue} is striking; the absorption spectrum is made up of innumerable absorption lines, each with their own strength, with line strengths varying by many orders of magnitude across the longwave spectrum. However, despite this fine-scale complexity, there is large-scale `structure' to the absorption coefficients. For example, H$_{2}$O line absorption exhibits multiple well-known `bands', such as the pure rotation band ($200<\tilde{\nu}<1000$ cm$^{-1}$), the vibration-rotation band ($1000<\tilde{\nu}<1450$ cm$^{-1}$), and the combination band ($1700<\tilde{\nu}<2500$ cm$^{-1}$). In these bands, line absorption either declines exponentially with wavenumber (rotation, combination), or increases exponentially (vibration-rotation). Also notable are the H$_{2}$O `window' regions, where line absorption is very weak and where continuum absorption dominates over line absorption. The most notable of these is the window region centered around 1000 cm$^{-1}$. CO$_{2}$ also has a complicated spectral structure, organized into a series of exponential increases and decays which appear as `triangles' on the logarithmic axis used in Figure \ref{fig-kappaTrue}. Of particular interest for Earth-like climates is the prominent absorption feature around 667 cm$^{-1}$ (sometimes referred to as the `15 $\mu$m band'), which is close to the peak of the Planck radiation curve for an Earth-like blackbody \cite{wordsworth2024fermi}. CO$_{2}$ also has opacity in other bands, but the `15 $\mu$m band' is the source of the vast majority of radiative forcing at modern atmospheric concentrations \cite{mlynczak2016spectroscopic, romps2022forcing}. 

% Talk about when is it opaque or transparent??? e.g. Romps et al 2022?

Efficiently representing this spectral structure is a key problem in parameterizing radiative transfer for GCMs. Correlated-k models typically achieve this by splitting the spectrum into a small number of bands, performing a number of `pseudo-monochromatic' calculations within these bands. In practice, lookup tables are generated for a number of different temperatures and pressures and the results are linearly interpolated to the GCM temperature and pressure at runtime. On the other hand, gray models neglect this spectral structure entirely and instead use absorption coefficients (or more commonly, optical depths) which are independent of wavenumber \cite{Frierson06a, jeevanjee2020simple}. These differences neatly reflect the `clarity-accuracy' trade-off made by these two approaches (Fig. \ref{fig-introschematic}): correlated-k methods are more accurate but rely heavily on empirical representations of the underlying spectroscopy, whereas gray schemes represent the spectroscopy simply, but inaccurately. 

In contrast to these two approaches, the key principle of simple spectral modeling is that we can neglect the fine-scale spectroscopic details of $\kappa_{\tilde{\nu}}$ and instead represent its `large-scale' spectral variations using analytical fits. It is not obvious \textit{a priori} that this should work, but we will show that solely representing the `large-scale' structure of $\kappa_{\tilde{\nu}}$ is sufficient to capture many key features necessary for an accurate climate simulation. Using analytical fits to represent $\kappa_{\tilde{\nu}}$ in a GCM's radiative transfer scheme builds off of previous work \cite<e.g.>{wilson2012simple,jeevanjee2020simple,koll23,spaulding2025clear,cohen2025spectroscopic} which have used similar approximations to develop pencil-and-paper models of radiative cooling and climate feedbacks, and early work by \citeA{crisp1986approximate} who used an analytic approximation to CO$_{2}$'s absorption spectrum when developing an early radiative transfer parameterization.

\begin{figure}
 \noindent\includegraphics[width=\textwidth]{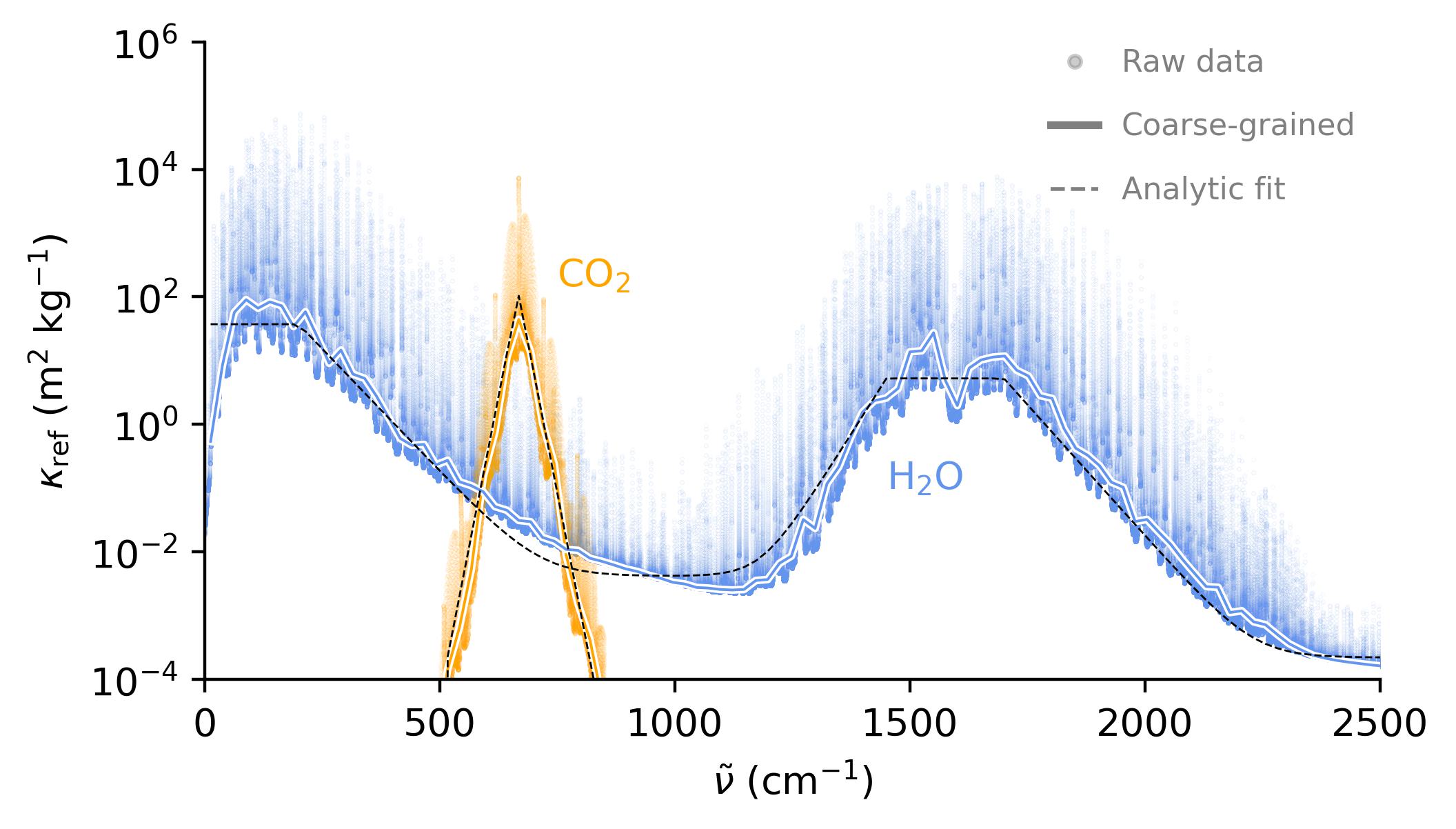}
\caption{Reference mass absorption coefficients (m$^{2}$ kg$^{-1}$) for H$_{2}$O line and continuum absorption (light blue) and CO$_{2}$ line absorption (orange) at $\left(T_{\mathrm{ref}}, \,  p_{\mathrm{ref}}\right)$ = (260K, 500hPa) and 100\% relative humidity calculated using PyRADS \cite{koll18}. We only consider CO$_{2}$ absorption in the 500-850cm$^{-1}$ range, but this could easily be extended if the user desires. The coarse-grained absorption coefficients for CO$_{2}$ and H$_{2}$O are shown in the thick lines, and the dashed black lines show the piece-wise analytic fits to these coarse-grained absorption coefficients.}
\label{fig-kappaFits}
\end{figure}

We begin by outlining the analytical fits we use to parameterize line absorption from H$_{2}$O and CO$_{2}$, and continuum absorption from H$_{2}$O. We perform these fits at the reference conditions outlined above, then use simple scalings to extend the results to different temperatures and pressures.

For line absorption from H$_{2}$O, we approximate the reference absorption coefficients as a piecewise function of wavenumber, $\tilde{\nu}$,

\begin{equation}
\kappa^{\mathrm{ref}}_{\mathrm{H}_{2}\mathrm{O, line}} = 
\begin{cases}
\kappa_{\mathrm{rot}} & \text{if }  \tilde{\nu} \in [10 \,\mathrm{cm}^{-1}, 200\,\mathrm{cm}^{-1}) \\
\kappa_{\mathrm{rot}} \exp\left(-\frac{\tilde{\nu}-200}{l_{\mathrm{rot}}}\right) & \text{if }  \tilde{\nu} \in [200 \,\mathrm{cm}^{-1}, 1000\,\mathrm{cm}^{-1}) \\
\kappa_{\mathrm{vr}} \exp\left(-\frac{1450-\tilde{\nu}}{l_{\mathrm{vr,1}}}\right) & \text{if } \tilde{\nu} \in [1000 \,\mathrm{cm}^{-1}, 1450\,\mathrm{cm}^{-1}) \\
\kappa_{\mathrm{vr}} & \text{if } \tilde{\nu} \in [1450 \,\mathrm{cm}^{-1}, 1700\,\mathrm{cm}^{-1}) \\
\kappa_{\mathrm{vr}} \exp\left(-\frac{\tilde{\nu}-1700}{l_{\mathrm{vr,2}}}\right) & \text{if } \tilde{\nu} \in [1700 \,\mathrm{cm}^{-1}, 2500\,\mathrm{cm}^{-1}] \\
\end{cases}
\label{eqn-h2oline}
\end{equation}

\noindent where the `rot' and `vr' subscripts are shorthand for the `rotation' and `vibration–rotation' bands of H$_{2}$O. 

For CO$_{2}$, we approximate the reference absorption coefficients as

\begin{equation}
    \kappa^{\mathrm{ref}}_{\mathrm{CO}_{2}} = \kappa_{\mathrm{CO_{2}}} \exp\left(-\frac{|\tilde{\nu}-\tilde{\nu}_{\mathrm{CO_{2}}}|}{l_{\mathrm{CO_{2}}}}\right) \quad \text{if } \tilde{\nu} \in [500 \,\mathrm{cm}^{-1}, 850\,\mathrm{cm}^{-1}].
\label{eqn-co2line}
\end{equation}

This formulation (Eq. \ref{eqn-co2line}) focuses on the strongly-absorbing `bending mode' of CO$_{2}$ and ignores the absorption from secondary CO$_{2}$ bands. We find this to be a reasonable approximation, but other modes of CO$_{2}$ absorption could easily be added back into the SSM by defining a piece-wise $\kappa^{\mathrm{ref}}_{\mathrm{CO}_{2}}$ as we have done for H$_{2}$O. 

We model the H$_{2}$O continuum absorption as two separate, grey, absorbers:

\begin{equation}
\kappa^{\mathrm{ref}}_{\mathrm{H}_{2}\mathrm{O, cnt}} = 
\begin{cases}
\kappa_{\mathrm{cnt,}1} & \text{if }  \tilde{\nu} \in [10 \,\mathrm{cm}^{-1}, 1700\,\mathrm{cm}^{-1}) \\
\kappa_{\mathrm{cnt,}2} & \text{if } \tilde{\nu} \in [1700 \,\mathrm{cm}^{-1}, 2500\,\mathrm{cm}^{-1}].\\
\end{cases}
\label{eqn-h2ocontinuum}
\end{equation}

In total, our model of H$_{2}$O and CO$_{2}$ spectroscopy has 10 parameters (Table \ref{table-values}). To fit these parameters, we first group the absorption coefficients into 25 cm$^{-1}$ bins and calculate the median absorption coefficient in each bin. We take the median because the average transmission across a spectral band tends to be dominated by the most optically-thin frequencies \cite{Pierrehumbert10, koll23}. We then use \url{scipy.optimize.curve_fit} to fit Equations (\ref{eqn-h2oline}), (\ref{eqn-co2line}), and (\ref{eqn-h2ocontinuum}) to these coarse-grained absorption coefficients \cite{virtanen2020scipy}. We first fit Eq. (\ref{eqn-h2oline}) to the line absorption for H$_{2}$O only, then use these parameters to fit our continuum model (Eq. \ref{eqn-h2ocontinuum}) to the sum of line and continuum absorption. We fit Eq. (\ref{eqn-co2line}) to the line absorption for CO$_{2}$ in isolation. The results of this fitting procedure are shown in Table \ref{table-values}. The results are somewhat sensitive to the details of the fitting procedure, but not to the extent that they qualitatively affect the accuracy of the resultant GCM simulation. 

\begin{table}
\caption{Spectroscopic parameters in the Simple Spectral Model (SSM). These are the reference values, taken at ($T_{\mathrm{ref}}$, $p_{\mathrm{ref}}$, RH$_{\mathrm{ref}}$) = $(260\mathrm{K}, 500\mathrm{hPa}, 100\%)$. }
\centering
\begin{tabular}{l c r}
\hline
 Parameter  & Value & Unit  \\
\hline
  $\kappa_{\mathrm{rot}}$  & 37 &  $\mathrm{m}^{2}  \,\mathrm{kg}^{-1}$ \\
  $\kappa_{\mathrm{vr}}$  & 5  &  $\mathrm{m}^{2}  \,\mathrm{kg}^{-1}$\\
  $l_{\mathrm{rot}}$  & 56 & cm$^{-1}$  \\
  $l_{\mathrm{vr,1}}$  & 37  & cm$^{-1}$ \\
  $l_{\mathrm{vr,2}}$  & 52 & cm$^{-1}$  \\
  $\kappa_{\mathrm{cnt,}1}$ & 0.004 &   $\mathrm{m}^{2}  \,\mathrm{kg}^{-1}$\\
  $\kappa_{\mathrm{cnt,}2}$  & 0.0002  &  $\mathrm{m}^{2}  \,\mathrm{kg}^{-1}$\\
  $\kappa_{\mathrm{CO}_{2}}$  & 110 &  $\mathrm{m}^{2} \, \mathrm{kg}^{-1}$ \\
  $\tilde{\nu}_{\mathrm{CO}_{2}}$  & 667  & cm$^{-1}$ \\
  $l_{\mathrm{CO}_{2}}$  & 12   & cm$^{-1}$ \\
\hline
%\multicolumn{2}{l}{$^{a}$Footnote text here.}
\end{tabular}
\label{table-values}
\end{table}
Figure \ref{fig-kappaFits} shows the reference absorption coefficients for CO$_{2}$ (in the region of the 15 $\mu$m band) and H$_{2}$O (the sum of line and continuum absorption), the coarse-grained absorption coefficients, and the piece-wise analytic fits from Equations (\ref{eqn-h2oline}), (\ref{eqn-co2line}), and (\ref{eqn-h2ocontinuum}). Overall, the piece-wise approximations capture the shape of the absorption coefficients fairly well. There is a slight underestimation of H$_{2}$O absorption at the edge of the main window region, but this is a region where absorption from CO$_{2}$ dominates and so we still consider the fits in Figure \ref{fig-kappaFits} to be adequate. 

To extend these fitted absorption coefficients beyond reference conditions we multiply $\kappa^{\mathrm{ref}}_{\mathrm{CO}_{2}}$ and $\kappa^{\mathrm{ref}}_{\mathrm{H}_{2}\mathrm{O, line}}$ by a pressure-scaling factor, $p/p_{\mathrm{ref}}$, which accounts for `foreign-broadening' of spectral lines due to collisions with other molecules \cite{Pierrehumbert10}. We neglect any temperature-scaling of the reference absorption coefficients for line absorption, following \citeA{koll23}. The H$_{2}$O continuum experiences `self-broadening' and thus scales with the water vapor pressure, $p_{v}/p_{v,\mathrm{ref}}$ \cite{shine2012water, jeevanjee2021analytical}. We also include an exponential temperature-dependence of the continuum, $e^{\sigma(T_{\mathrm{ref}}-T)}$, following \citeA{mlawer97}, with $\sigma=0.02$ K$^{-1}$.

Including these pressure- and temperature-scalings, our analytical representations of H$_{2}$O line absorption, H$_{2}$O continuum absorption \& CO$_{2}$ line absorption are:

\begin{equation}
    \kappa_{\mathrm{H}_{2}\mathrm{O,line}}\left(\tilde{\nu}, p\right) =  \left(\frac{p}{p_{\mathrm{ref}}}\right)  \, \kappa^{\mathrm{ref}}_{\mathrm{H}_{2}\mathrm{O,line}}(\tilde{\nu}),
\label{eqn-h2oline_broadened}
\end{equation}
\begin{equation}
    \kappa_{\mathrm{H}_{2}\mathrm{O,cnt}}\left(\tilde{\nu}, p, T\right) =  \left(\frac{p_{v}}{p_{\mathrm{v,ref}}}\right)  \, e^{\sigma(T_{\mathrm{ref}}-T)} \, \kappa^{\mathrm{ref}}_{\mathrm{H}_{2}\mathrm{O,cnt}}(\tilde{\nu}),
\label{eqn-h2ocnt_broadened}
\end{equation}
and,
\begin{equation}
    \kappa_{\mathrm{CO}_{2}}\left(\tilde{\nu}, p\right) =  \left(\frac{p}{p_{\mathrm{ref}}}\right)  \, \kappa^{\mathrm{ref}}_{\mathrm{CO}_{2}}(\tilde{\nu}).
\label{eqn-co2line_broadened}
\end{equation}

We have now constructed our SSM. With Equations (\ref{eqn-twostreamLWup}-\ref{eqn-opticaldepth}) and (\ref{eqn-h2oline_broadened}-\ref{eqn-co2line_broadened}) in hand, and the parameters in Table \ref{table-values}, we can simulate longwave fluxes upwards and downwards through the atmosphere given profiles of temperature and humidity. We note that our SSM differs from that of \citeA{jeevanjee2020simple} in that they invoke the `cooling-to-space' approximation, whereas we numerically integrate the radiative transfer equations (Eq. \ref{eqn-twostreamLWup}-\ref{eqn-opticaldepth}). 

As a sanity check, in Appendix A we implement our SSM in a single-column model and show it can reproduce the state-dependence of the longwave feedback parameter found by previous studies \cite{koll23, stevens2023colorful}. We note that the SSM's detailed spectral output also allows us to easily understand the radiative processes involved, in contrast to broadband radiation schemes. Satisfied that our SSM can reproduce previous single-column model results, next we implement our SSM scheme in an idealized GCM and compare the resulting zonal-mean climate to that obtained with a gray scheme and RRTMG.

\section{Idealized GCM simulations}

\subsection{Model setup}

The idealized aquaplanet GCM we use is a configuration of Isca \cite{vallis18} that follows closely \citeA{Frierson06a} and \citeA{OGorman08b}. The representation of physical processes in the atmosphere is idealized in a variety of ways: the `microphysics' is a saturation adjustment to $100\%$ relative humidity using an idealized formulation of the Clausius--Clapeyron relation, condensed water instantaneously falls to the surface as precipitation, a simplified Betts-Miller scheme relaxes convectively unstable profiles toward a moist adiabat with a $2$-hour timescale \cite{frierson07b}, and there is no representation of sea-ice in the model. The model also has no representation of clouds, which is appropriate as our focus is on the climatic impacts of differences in clear-sky radiative transfer. All simulations are performed at T42 spectral truncation ($\approx 2.8^\circ$ horizontal resolution) with $50$ vertical levels in the model's sigma (pressure divided by surface pressure) coordinate and are integrated for 6000 days. The analysis is conducted over the final 3000 days once the simulations have reached a steady-state. 

We run the model with a hierarchy of clear-sky longwave radiative transfer schemes. Specifically, we run Isca with our longwave simple spectral model (SSM) and compare the results to simulations using either gray radiation or a correlated-k scheme. As a representative gray scheme we use the \citeA{Frierson06a} scheme, in which the (gray) optical depth is a prescribed function of latitude and pressure. The correlated-k scheme we use is RRTMG \cite{mlawer97, iacono2008radiative}. We do not include cloud radiative effects or ozone in any of our simulations, and run the SSM and RRTMG with 300ppmv of CO$_{2}$.

At each GCM timestep, we solve the SSM equations at 41 equally-spaced wavenumbers between 10 cm$^{-1}$ and 2500 cm$^{-1}$, then spectrally-integrate to get the broadband heating rates. Results are qualitatively similar if we solve the equations either at 20 or 100 equally-spaced wavenumber points, or extend the upper-bound of wavenumber space to 3000 cm$^{-1}$. 

To aid comparison between the hierarchy of different longwave radiation schemes, the treatment of shortwave radiation is the same between the three sets of experiments. The insolation is a time-independent distribution that approximates a perpetual annual-mean using a second Legendre polynomial specification and there is no atmospheric absorption of solar radiation (thus the shortwave radiative heating rate is zero at all points in the atmosphere). There is no diurnal or seasonal cycle in the model.

The focus of this paper is not on the `efficiency-accuracy' trade-off, which has been discussed in detail by other authors \cite<e.g.,>{pincus2019balancing}, but we note in passing that running the GCM with our SSM is about four times faster than running the GCM with RRTMG\_LW. 

\subsection{Control climate}   

\begin{figure}
 \noindent\includegraphics[width=\textwidth]{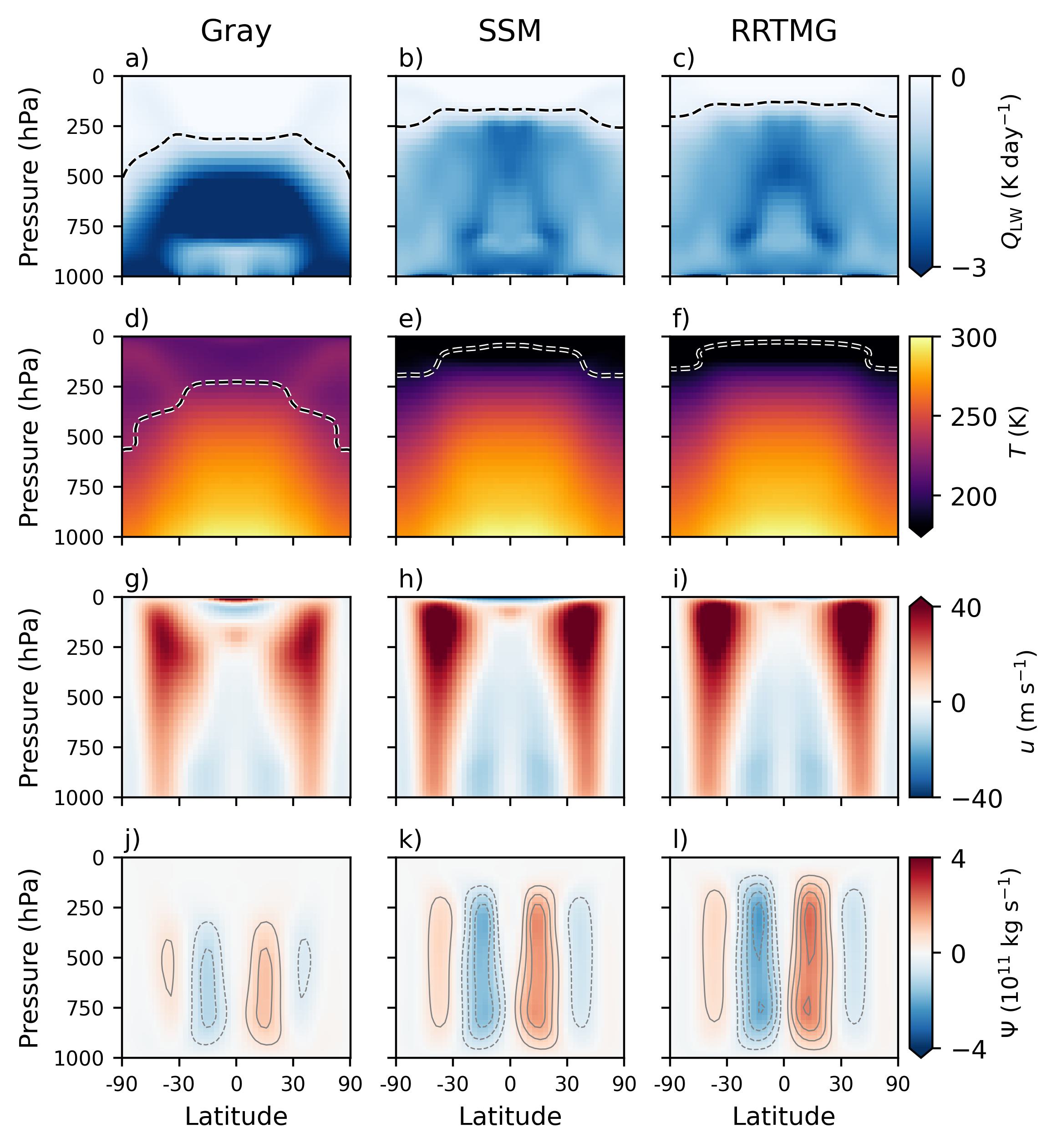}
\caption{Zonal-mean (a-c) longwave radiative heating rate, (d-f) temperature, (g-i) zonal wind, and (j-l)  mean meridional streamfunction for GCM simulations using (left column) gray radiation, (middle column) the simple spectral model, and (right column) RRTMG\_LW. Contours in (a-c) and (d-f) are the radiative and lapse-rate tropopause respectively, diagnosed as the -0.3 K day$^{-1}$ and -5 K km$^{-1}$ contours \cite{mckim2025water}. Contours in (j-l) are plotted at 5x10$^{10}$ kg s$^{-1}$ intervals, with negative contours dashed and the zero contour omitted. Horizontal axis is the sine of latitude.}
\label{fig-gcmControl}
\end{figure}

In Figure \ref{fig-gcmControl} we compare the time-averaged, zonal-mean climates of the idealized GCM coupled to the three longwave radiation schemes. Because the GCM simulations are run with Earth-like SSTs and thus are within RRTMG's range of validity, we will treat the RRTMG results as a benchmark against which to compare the results using gray scheme and the SSM.

The longwave heating rate is negative throughout the troposphere, indicating that the atmosphere is cooled by longwave emission (Fig. \ref{fig-gcmControl}a-c). However, while RRTMG and SSM generate a radiative cooling profile which is around $\approx-2$K day$^{-1}$ throughout the atmosphere (in line with observations \cite{jeevanjee2020simple}), the gray model produces a pronounced `bulge' of radiative cooling in the upper troposphere whose magnitude is much larger (peaking at around -5 K day$^{-1}$). These differences in the radiative cooling profiles can be understood simply through the `unit optical depth' approximation, which states that the radiative cooling (specifically the cooling-to-space), maximizes at the level where $\tau=1$ \cite<e.g.,>{petty2006first, jeevanjee2020cooling}. For gray schemes, the optical depth has no wavenumber-dependence, and so there is a single `$\tau=1$' level where all of the radiative cooling becomes concentrated, whereas for the SSM and RRTM there are a range of `$\tau=1$' levels throughout the atmosphere (each at a different wavenumber), yielding a more diffuse radiative cooling profile. There are small differences between the SSM and RRTMG configurations, but the SSM nevertheless produces a significantly more accurate simulation of longwave cooling profiles—that is, closer to RRTMG—compared to the gray model.

%Panels (a-c) of Figure \ref{fig-gcmControl} show the zonal-mean longwave radiative heating rates for the three configurations. The longwave heating rate is negative throughout the troposphere, indicating that the atmosphere is cooled by longwave emission. However, while RRTMG and SSM generate a radiative cooling profile which is around $\approx-2$K day$^{-1}$ throughout the atmosphere (in line with observations \cite{jeevanjee2020simple}), the gray model produces a pronounced `bulge' of radiative cooling in the upper troposphere whose magnitude is much larger (peaking at around -5 K day$^{-1}$). These differences in the radiative cooling profiles can be understood simply through the `unit optical depth' approximation, which states that the radiative cooling (specifically the cooling-to-space), maximizes at the level where $\tau=1$ \cite<e.g.,>{petty2006first, jeevanjee2020cooling}. For gray schemes, the optical depth has no wavenumber-dependence, and so there is a single `$\tau=1$' level where all of the radiative cooling becomes concentrated, whereas for the SSM and RRTM there are a range of `$\tau=1$' levels throughout the atmosphere (each at a different wavenumber), yielding a more diffuse radiative cooling profile. There are small differences between the SSM and RRTMG configurations, but the SSM nevertheless produces a significantly more accurate simulation of longwave cooling profiles—that is, closer to RRTMG—compared to the gray model.

The pressure at which the radiative cooling approaches zero denotes the `radiative tropopause', which separates the dynamically-active troposphere (where radiative cooling is balanced by heating from convection and eddies) from the largely quiescent stratosphere (which is approximately in radiative equilibrium). The radiative tropopause, diagnosed as the upper-tropospheric -0.3 K day$^{-1}$ contour, is significantly lower in the atmosphere in the gray model compared to the SSM or RRTMG configurations, and also has a more pronounced latitudinal structure (Fig. \ref{fig-gcmControl}(a-c)). 

The most notable differences in zonal-mean temperature between the gray scheme and the SSM/RRTMG schemes are in the upper-troposphere and stratosphere (Fig. \ref{fig-gcmControl}d-f). For example, the lapse-rate tropopause, diagnosed at the level where the lapse-rate first falls below -5 K km $^{-1}$ following \citeA{mckim2025water}, is lower down in the atmosphere in the gray configuration compared to RRTMG or indeed the SSM. This is consistent with the radiative tropopause. SSM exhibits a small `too-low' bias in both the lapse-rate and radiative tropopause (the tropically-averaged radiative tropopause pressure is 322 hPa for gray, 161 hPa for SSM, and 134 hPa for RRTMG), which is consistent with previous work by \citeA{mckim2025water} who showed that the temperature (and height) of the tropopause is tied to water vapor's maximum absorption coefficient, which is underestimated by our piece-wise analytic fits (Fig. \ref{fig-kappaFits}).

We also note that the stratospheric temperature is much warmer in the gray configuration compared to RRTMG or SSM. Remember that neither the SSM or RRTMG have a representation of shortwave absorption from ozone, which warms the stratosphere \cite{manabe67}. The gray model exhibits temperature inversions in the extratropical stratosphere at around 200 hPa, where temperatures again increase with height. This results in a reversed meridional temperature gradient in the stratosphere of the gray model, where tropical temperatures are colder than extratropical/polar temperatures. Many of these features of the gray model were previously noted by \citeA{tan19}. 

Because of thermal wind balance, the different temperature structures in the three models result in different zonal wind profiles (Fig. \ref{fig-gcmControl}(g-i)). All three configurations produce a deep, eddy-driven jet which drives surface westerlies but the gray model produces a shallower and weaker jet compared to the models with RRTMG or SSM. The latitude of the eddy-driven jet, diagnosed as the latitude of maximum 850 hPa zonal-mean zonal wind \cite{kang2011interannual}, is about 41-42$^{\circ}$ for all three configurations. However, the subtropical jet latitude, diagnosed as the latitude of maximum zonal-mean zonal wind between 50 and 400 hPa with the 850 hPa wind removed \cite{waugh2018revisiting}, is 26$^{\circ}$ for the gray model compared to 34$^{\circ}$ for the SSM and 36$^{\circ}$ for RRTMG. This `split jet' in the gray model is consistent with the fact that the tropopause `break', the region of maximum baroclinicity where the tropopause abruptly shifts from its tropical to extratropical altitude, is about 10$^{\circ}$ further equatorward in the gray model compared to either the RRTMG or SSM configurations. The subtropical jet is also weaker in the gray model compared to RRTMG, consistent with the idea that the thermal wind relation is integrated over a shallower troposphere in the gray model \cite{tan19}.

In Fig. \ref{fig-gcmControl}j-l we show the mean meridional streamfunction, calculated as
\begin{equation}
    \psi(p, \phi) = \frac{2\pi a \cos(\phi)}{g} \int_{p}^{0}[v]dp,
\end{equation}
where $p$ is the pressure, $\phi$ is the latitude, $a=6,371$km is the mean radius of Earth, $g=9.81$m s$^{-1}$ is the acceleration due to gravity, $v$ is the meridional wind, and square brackets indicate the zonal-mean. All three radiation schemes produce Hadley cells of similar latitudinal width. However, the gray scheme exhibits a weaker and shallower Hadley cell compared to the SSM and RRTMG, consistent with its more equatorward jet structure and lower tropopause height.

\subsection{Response to uniform warming}

\begin{figure}
 \noindent\includegraphics[width=\textwidth]{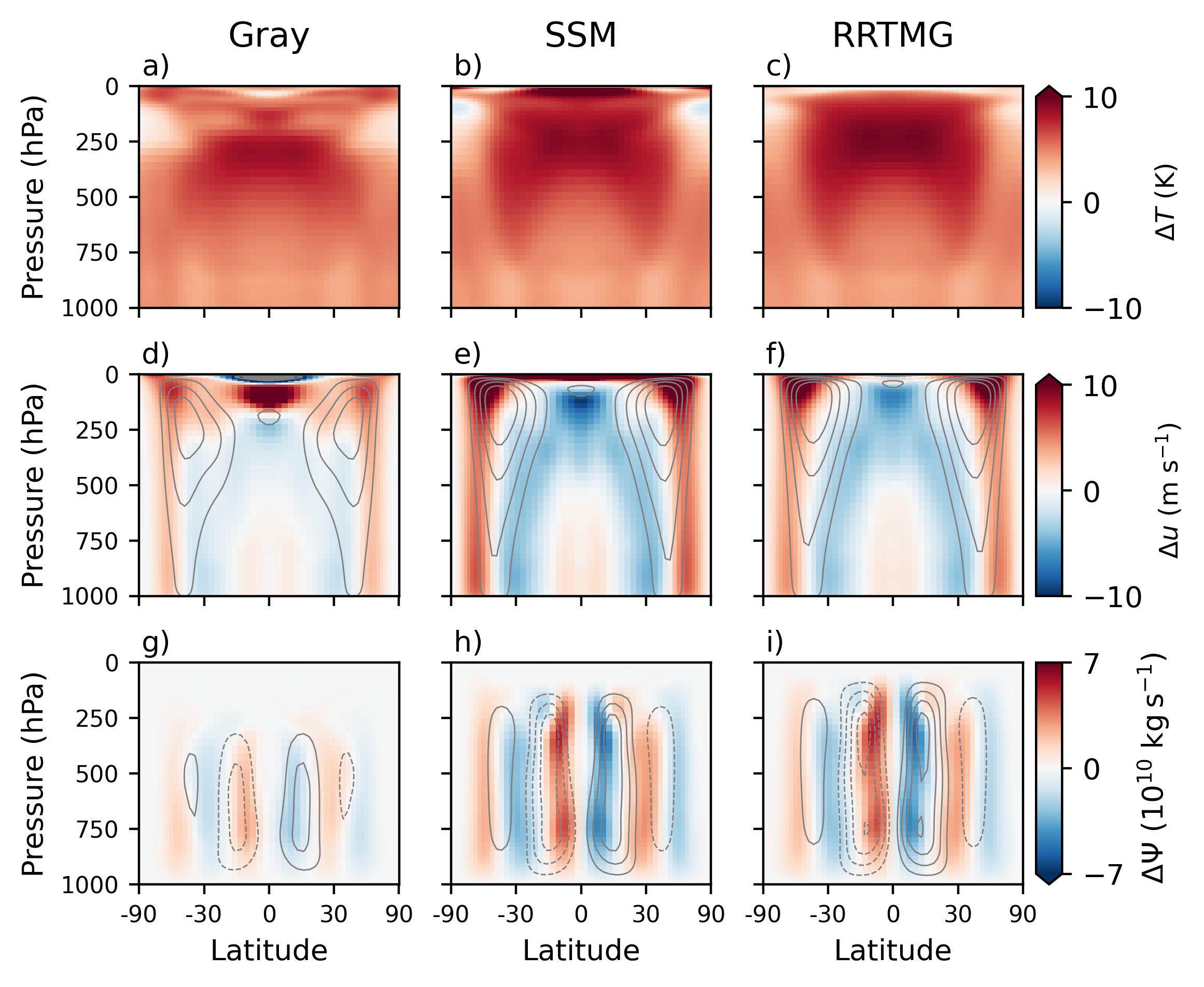}
\caption{Zonal-mean changes in (a-c) temperature, (d-f) zonal wind, and (g-i) mean meridional streamfunction in response to uniform warming for the GCM with gray radiation (left column), the SSM (middle column), and RRTMG\_LW (right column). Contours in (d-f) represent the 10, 20, 30, 40 m s$^{-1}$ winds in the control run. Contours in (g-i) are the same as (j-l) of Figure \ref{fig-gcmControl}. Horizontal axis is the sine of latitude.}
\label{fig-gcmPerturbed}
\end{figure}

The SSM's idealized representation of spectroscopy also captures key features of the coupling between radiation and dynamics under warming, as shown by Figure \ref{fig-gcmPerturbed} where we plot the change in zonal-mean temperature (a-c), zonal wind (d-f), and mean meridional streamfunction (g-i) in response to a uniform 4K warming of the SSTs.

All three radiation schemes exhibit an amplification of temperature change with altitude in the tropics (Fig. \ref{fig-gcmPerturbed}a-c), consistent with the expectation from moist adiabatic physics. However, the magnitude and vertical structure of tropical upper-tropospheric amplification differs substantially between the gray model and RRTMG/SSM. To quantify this, we calculate an `amplification factor' as the ratio of tropically-averaged temperature change at 200hPa compared to the change at 1000hPa. The amplification factor is 1.68 for gray radiation, compared to 2.37 for SSM and 2.35 for RRTMG. Visually, it is also apparent that the gray model produces weaker warming of the subtropical troposphere compared to RRTMG or SSM. 

Under warming we also expect the tropopause to move to higher altitudes \cite{santer2003behavior,vallis15, mckim2025water}, and this expectation is borne out in all three radiation schemes. Quantitatively, the pressure of the tropically-averaged radiative tropopause changes from 322 hPa to 295 hPa for gray, 161 hPa to 140 hPa for SSM, and 134 hPa to 108 hPa for RRTMG. The magnitude of the vertical shift is similar between all schemes, and consistent with the simple scaling of about $6$ hPa K$^{-1}$ proposed by \citeA{match2021large}. 

All schemes exhibit a poleward shift of the subtropical and eddy-driven jet by $\approx 1-2$ degrees (Fig. \ref{fig-gcmPerturbed}d-f). The poleward shift of the eddy-driven jet is visually evidenced by the contrasting zonal-wind changes on either side of the surface westerlies. All schemes also simulate an intensification of the surface easterlies on the equator, and a general weakening of westerlies on the equatorward side of the jet. However, the gray model produces changes in the upper-troposphere which differ from RRTMG or SSM. Most notably, while all schemes produce a super-rotating equatorial jet in their control climates which shifts upwards with warming, the gray model is the only configuration which features a strong intensification of the super-rotation with warming. Additionally, while RRTMG and SSM produce a weakening of the upper-tropospheric flow on the equatorward side of the jet, the gray model produces a strengthening on the equatorward side of the upper-tropospheric jet core. This is possibly related to an intensification of the subtropical jet, which is still `split' from the eddy-driven jet in the gray model in response to uniform warming.

These gray radiation results, in which the subtropical and eddy-driven jets shift poleward with warming, appear contrary to previous work which has reported that the jet shifted \textit{equatorward} with warming in gray radiation models \cite{Schneider10, dwyer17, lachmy2018connecting, tan19}. However, the equatorward shift seen in previous studies is sensitive to the parameters used in the gray scheme, particularly the `$f_{l}$' parameter \cite{Frierson06a, OGorman08b}. We set $f_{l}=0.1$ in this work, following the original implementation of \citeA{Frierson06a}, but \citeA{davis2022eddy} showed that setting $f_{l}=0.2$ as in \citeA{OGorman08b} altered the climatological jet structure substantially. These minor changes to the gray scheme also affect the response of the jet latitude to warming in slab ocean configurations (Zhihong Tan, personal communication). Given that both the climatological circulation structure and its response to warming are so sensitive to such ad hoc parameter choices in the gray model, this further underlines the importance of developing idealized, but physically-motivated radiation schemes like the SSM whose parameters represent actual physical quantities and can therefore be constrained by spectroscopic observations.

The Hadley Cell changes (Fig. \ref{fig-gcmPerturbed}g-i) are also much better represented when using SSM than gray radiation. All three schemes yield a weakening and polewards expansion of the Hadley Cell, but these changes are muted in the gray model compared to RRTMG or the SSM. The Hadley Cell width, diagnosed as the latitude closest to the equator where the streamfunction (vertically averaged with mass weighting between 700 and 300 hPa) is zero \cite{byrne18}, increases by 1.5$^{\circ}$ in the gray model, compared to roughly 2.1$^{\circ}$ in RRTMG and SSM. Furthermore, the Hadley Cell strength, diagnosed by vertically averaging the streamfunction magnitude with mass weighting between 700 and 300 hPa and then taking the meridional average between the equator and the cell edge, decreases by around 1.5x10$^{9}$ kg s$^{-1}$ in the gray model, compared to much larger decreases of 6.8x10$^{9}$ kg s$^{-1}$ and 8.4x10$^{9}$ kg s$^{-1}$ in the SSM and RRTMG, respectively. We also note that the vertical structure of Hadley Cell changes is different between the gray model and SSM/RRTMG. RRTMG and SSM have streamfunction changes which maximize in the upper-troposphere, whereas the gray model has streamfunction changes which maximize in the lower-troposphere, and in fact the gray model exhibits \textit{increases} in the cell-mean Hadley Cell strength in the upper troposphere.

These GCM results show that our SSM is able to accurately represent the large-scale structure of the atmosphere and its response to climate change compared to RRTMG, which is validated in this range of temperatures and thus can be taken as a benchmark. This is in stark contrast to the gray model, which distorts many aspects of the large-scale atmospheric circulation and its response to warming. This demonstrates that a simplified representation of H$_{2}$O and CO$_{2}$ spectroscopy is sufficient to capture the key features of radiation-dynamics coupling in both current and future Earth-like climates.

\section{Conclusions}

We began this paper by noting a conspicuous gap in the hierarchy of radiative transfer schemes used in climate modeling; there are accurate, but somewhat opaque, correlated-k schemes and conceptually simple, but inaccurate, gray schemes. To bridge this `clarity-accuracy' gap, we have introduced a new clear-sky, longwave radiative transfer scheme—the Simple Spectral Model (SSM). Previous work has shown that significant theoretical insight into radiative transfer problems can be gained by representing the spectral structure of absorption coefficients using piece-wise analytic fits \cite<e.g.,>{jeevanjee2020simple,koll23,fildier2023moisture, mckim2025water}. Our SSM leverages this insight to build a clear-sky, longwave radiative transfer scheme suitable for idealized climate models which is both easily intelligible and reasonably accurate.

When implemented in an idealized GCM, the SSM produces zonal-mean climates that are much closer to those of the benchmark correlated-k scheme RRTMG than to the widely-used gray radiation scheme of \citeA{Frierson06a}. Notable improvements are found in the representation of radiative cooling profiles, tropopause structure, atmospheric jets, and the Hadley Cell both in the control climate and in the response to uniform warming. 

Correlated-k schemes are usually based on pre-calibrated lookup tables and thus are susceptible to errors when pushed outside of their training data \cite<e.g.,>{popp2016transition, kluft2021temperature}. In contrast to this, the SSM is equation-based and thus maintains physical consistency across a wide range of climates (Appendix A). The SSM also provides detailed spectral output, in contrast to broadband schemes, which greatly aids physical interpretation of results.

Future work could extend the SSM framework to include shortwave absorption by water vapor, in a similar manner to \citeA{roemer2025think}. This would allow the SSM to reproduce the negative hydrological sensitivity in hothouse climates noted by \citeA{liu2024hydrologic}, and to better understand the role of shortwave absorption on the general circulation of the atmosphere \cite{kang2019collapse}. The SSM framework could also be extended to include simple representations of scattering and clouds \cite{Pierrehumbert10}. While we have focused on including spectral representations of H$_{2}$O and CO$_{2}$ this could easily be updated to include other species, allowing exploration of methane-rich climates such as Titan \cite{mitchell2016climate} or exoplanets with even more exotic atmospheres \cite{kempton2011atmospheric,alderson2023early}. 

We hope to have demonstrated that the SSM is a useful tool for idealized climate modeling and fills in an important gap in the `hierarchy' of radiative transfer models. We also envisage this SSM being useful in classroom settings by helping students form a clearer link between the textbook physics of radiative transfer and the simulation of radiative transfer in GCMs. More broadly, we hope this work contributes to closing the long-standing gap between simulation and understanding in climate modeling \cite{held05b}.

\appendix
\section{State-dependence of longwave feedbacks in the SSM}

To examine the behavior of the SSM (and the gray model, for comparison) across a wide range of climate states, we conducted single-column radiative-convective equilibrium simulations with Isca \cite{mckim2025water} over prescribed sea surface temperatures (SSTs) with a vertically-constant relative humidity of 70\%. Our focus here is on the state-dependence of longwave radiative feedbacks and how the schemes represent high-temperature regimes. 

The single-column model simulations were run with 200 vertical levels to ensure a well-resolved tropopause and stratosphere even in extremely warm climates. We prescribed SSTs ranging from 260K to 330K in 5K increments and computed the clear-sky longwave feedback parameter, $\lambda_{\mathrm{LW}}$, in each climate as the change in outgoing longwave radiation per unit surface temperature change. The SSM simulations were run both without CO$_{2}$, to isolate H$_{2}$O effects, and then with a constant 300ppm of CO$_{2}$. The gray radiation scheme, by contrast, has no explicit representation of different absorbing species \cite{Frierson06a}.

\begin{figure}[h!]
 \noindent\includegraphics[width=\textwidth]{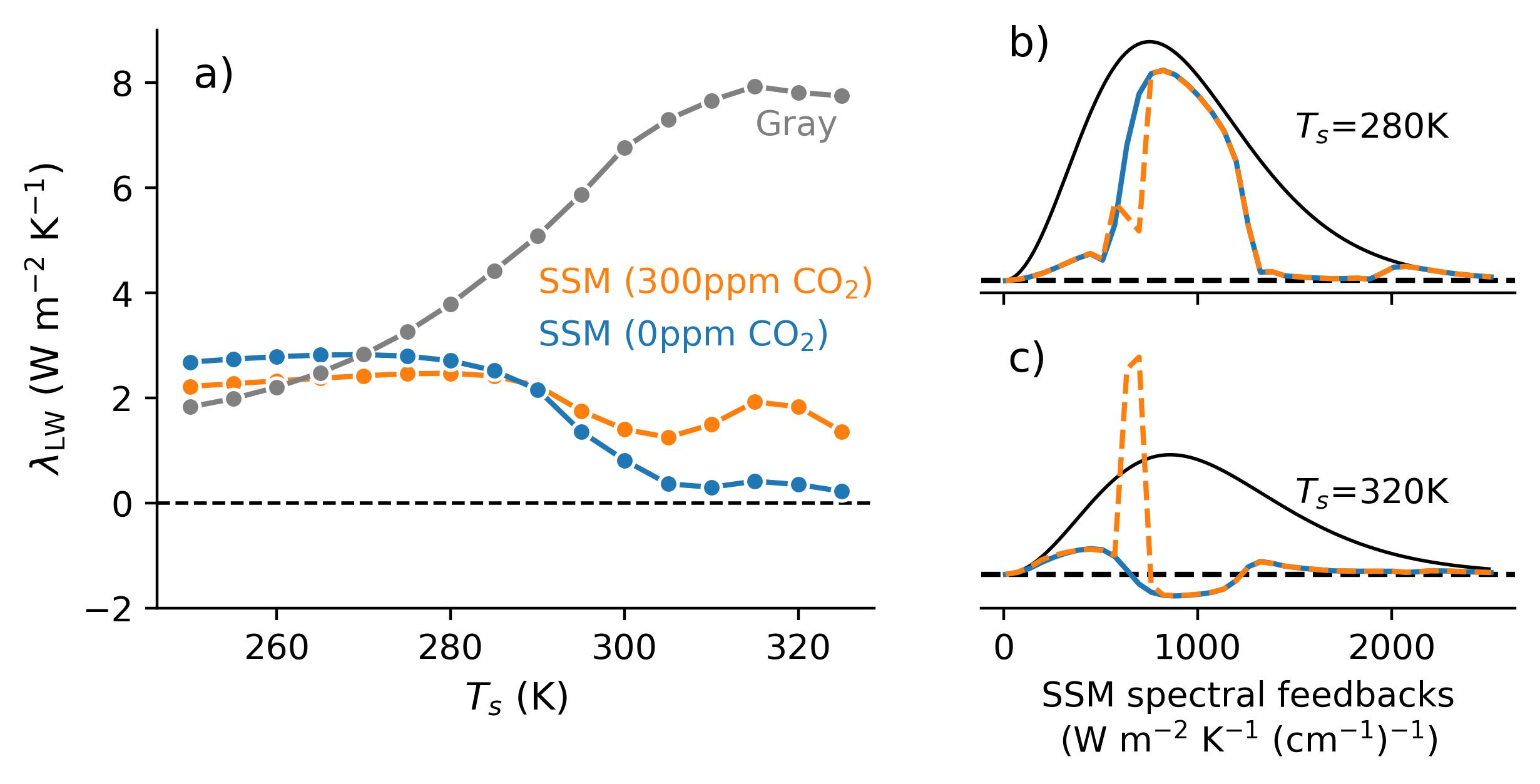}
\caption{(a) State-dependence of the clear-sky longwave feedback parameter, $\lambda_{\mathrm{LW}}$, in fixed-SST single-column runs using gray radiation, and the simple spectral model (SSM). The SSM simulations are conducted with 0ppm and 300ppm of CO$_{2}$. Panels (b) and (c) show the spectrally-resolved feedback parameter from the SSM for two baseline SSTs, blue solid curve is for 0ppm CO$_{2}$ and the orange dashed curve is for 300ppm CO$_{2}$. Black curves in (b) and (c) are the surface Planck feedback. All curves in (b) and (c) are normalized by the maximum value of the surface Planck feedback in that climate.}
\label{fig-widerange_lambda}
\end{figure}

Figure \ref{fig-widerange_lambda}a shows the clear-sky longwave feedback parameter, $\lambda_{\mathrm{LW}}$, as a function of baseline SST for the three sets of experiments. Both SSM set-ups have $\lambda_{\mathrm{LW}}\approx 2$ W m$^{-2}$ K$^{-1}$ for SSTs below 290K. This value of $\lambda_{\mathrm{LW}}$ arises because H$_{2}$O optical depths are an approximately fixed function of temperature and thus it is the optically-thin `window' regions which dominate changes in OLR with warming \cite{simpson1928olr,jeevanjee2021simpson}. Figure \ref{fig-widerange_lambda}b illustrates this by showing the spectrally-resolved change in OLR per degree of warming (i.e., the spectrally-resolved $\lambda_{\mathrm{LW}}$) in the SSM simulations for the 280K climate state. The vast majority of the extra OLR per degree warming is contributed by the window region centered around 1000 cm$^{-1}$. The orange dashed curve in Fig. \ref{fig-widerange_lambda}b shows the effect of adding a constant 300ppm of CO$_{2}$, which weakens the feedback at low temperatures by taking a `bite' out of the window region.

In warmer climates the exponential increase in water vapor with warming, coupled with the temperature-induced strengthening of the continuum due to self-broadening, causes the window regions to become optically-thick. When this happens, the longwave feedback in the window region rapidly weakens as shown by the SSM's spectral feedback parameter at 320K in Figure \ref{fig-widerange_lambda}c. This `closing of the water vapor window' is associated with a weakening of the total longwave feedback parameter (Fig. \ref{fig-widerange_lambda}a), which is particularly pronounced in the absence of CO$_{2}$, in line with line-by-line calculations \cite<e.g.>{goldblatt2013low,koll18,koll23}. We note that Fig. \ref{fig-widerange_lambda}c actually shows negative (destabilizing) feedbacks in the window region at 320K, which is because of the weakening of the moist lapse rate with warming which means that the temperature at which $\tau=1$ decreases slightly with warming \cite{jeevanjee2021simpson,koll23}.

Even in the H$_{2}$O-only case we expect that $\lambda_{\mathrm{LW}}$ will remain small and positive in the post-runaway state due to the influence of foreign pressure broadening \cite{feng2023stable}, an expectation borne out by the SSM in Fig. \ref{fig-widerange_lambda}a and the positive (stabilizing) feedbacks at wavenumbers below 700 cm$^{-1}$ in Fig. \ref{fig-widerange_lambda}c. However, what really `saves you' from the runaway greenhouse is not pressure broadening, but the presence of CO$_{2}$. Even preindustrial levels of CO$_{2}$ are sufficient to substantially strengthen $\lambda_{\mathrm{LW}}$ in warm climates compared to the 0ppm case, somewhat offsetting the effects of the closing of the water vapor window. The reason for this can be seen in Fig. \ref{fig-widerange_lambda}c, which shows that the presence of CO$_{2}$ causes additional outgoing longwave radiation to space from those wavenumbers where CO$_{2}$ is optically thick (just the 15 $\mu$m band, in our SSM). This is because CO$_{2}$ is well-mixed, and thus its $\tau=1$ levels are fixed functions of \textit{pressure} rather than temperature \cite{seeley2021h2o}. Warming of the troposphere thus leads to stabilizing feedbacks from CO$_{2}$. This is particularly evident near the center of the CO$_{2}$ band, which hits its $\tau=1$ level in the upper-troposphere and thus experiences greater warming per 1K increase in surface temperature due to moist adiabatic amplification of temperature changes in the upper-troposphere. 

Moving now to the gray model, we see that the gray radiation scheme exhibits an approximately monotonic increase in $\lambda_{\mathrm{LW}}$ with warming, reaching $\approx 8$ W m$^{-2}$ K$^{-1}$ at the warmest SSTs. This unphysical behavior stems from the scheme's inability to represent the spectral nature of atmospheric absorption and the existence of `window' regions. Although `windowed-grey' models exist \cite<e.g.,>{weaver1995deductions, geen2016effects}, these models assume assume that a fixed fraction of the longwave spectrum is covered by the window region and would thus miss the behavior in Figure \ref{fig-widerange_lambda}. One could imagine parameterizing this `window fraction' as a function of SST, but at that point the number of ad hoc parameters becomes unwieldy and, as discussed earlier (Section 4.2), the resulting climate states can be rather sensitive to these tunable parameters.

The SSM captures the state-dependence of $\lambda_{\mathrm{LW}}$ highlighted by previous studies \cite<e.g.>{koll23}. This includes the value of $\lambda_{\mathrm{LW}}\approx 2$ W m$^{-2}$ K$^{-1}$ at colder SSTs, the closing of the water vapor window at around 305K, and the CO$_{2}$-dependence of $\lambda_{\mathrm{LW}}$ at high SSTs. This demonstrates that the large-scale spectral structure captured by our piecewise fits (Fig. \ref{fig-kappaFits}) is sufficient to represent the essential physics of water vapor feedbacks, even in extreme climate states.

\acknowledgments

A.I.L. Williams acknowledges funding from the CIMES Postdoctoral Fellowship under award NA18OAR4320123 from the National Oceanic and Atmospheric Administration, U.S. Department of Commerce. \par 

%Thanks to Nadir Jeevanjee and Robert Pincus for many helpful discussions on life (and radiative transfer) over the years, and to Ilai Guendelman, Tim Merlis, and Neil Lewis for encouragement and helpful discussions. Thanks to Makayla and Cleo for moral support. \par

\section{Open Research}
Data supporting the conclusions, and FORTRAN90 code to implement the SSM in Isca, are archived at https://zenodo.org/records/16817101.

%% ------------------------------------------------------------------------ %%
%% References and Citations

%%%%%%%%%%%%%%%%%%%%%%%%%%%%%%%%%%%%%%%%%%%%%%%
%
% \bibliography{<name of your .bib file>} don't specify the file extension
%
% don't specify bibliographystyle
%%%%%%%%%%%%%%%%%%%%%%%%%%%%%%%%%%%%%%%%%%%%%%%

\bibliography{main}

\begin{thebibliography}{}

\bibitem [\protect \citeauthoryear {%
Alderson%
\ \protect \BOthers {.}}{%
Alderson%
\ \protect \BOthers {.}}{%
{\protect \APACyear {2023}}%
}]{%
alderson2023early}
\APACinsertmetastar {%
alderson2023early}%
\begin{APACrefauthors}%
Alderson, L.%
, Wakeford, H\BPBI R.%
, Alam, M\BPBI K.%
, Batalha, N\BPBI E.%
, Lothringer, J\BPBI D.%
, Adams~Redai, J.%
\BDBL {}others%
\end{APACrefauthors}%
\unskip\
\newblock
\APACrefYearMonthDay{2023}{}{}.
\newblock
{\BBOQ}\APACrefatitle {Early Release Science of the exoplanet WASP-39b with JWST NIRSpec G395H} {Early release science of the exoplanet wasp-39b with jwst nirspec g395h}.{\BBCQ}
\newblock
\APACjournalVolNumPages{Nature}{614}{7949}{664--669}.
\PrintBackRefs{\CurrentBib}

\bibitem [\protect \citeauthoryear {%
Armstrong%
}{%
Armstrong%
}{%
{\protect \APACyear {1968}}%
}]{%
armstrong1968theory}
\APACinsertmetastar {%
armstrong1968theory}%
\begin{APACrefauthors}%
Armstrong, B.%
\end{APACrefauthors}%
\unskip\
\newblock
\APACrefYearMonthDay{1968}{}{}.
\newblock
{\BBOQ}\APACrefatitle {Theory of the diffusivity factor for atmospheric radiation} {Theory of the diffusivity factor for atmospheric radiation}.{\BBCQ}
\newblock
\APACjournalVolNumPages{Journal of Quantitative Spectroscopy and Radiative Transfer}{8}{9}{1577--1599}.
\PrintBackRefs{\CurrentBib}

\bibitem [\protect \citeauthoryear {%
Byrne%
, Pendergrass%
, Rapp%
\BCBL {}\ \BBA {} Wodzicki%
}{%
Byrne%
\ \protect \BOthers {.}}{%
{\protect \APACyear {2018}}%
}]{%
byrne18}
\APACinsertmetastar {%
byrne18}%
\begin{APACrefauthors}%
Byrne, M\BPBI P.%
, Pendergrass, A\BPBI G.%
, Rapp, A\BPBI D.%
\BCBL {}\ \BBA {} Wodzicki, K\BPBI R.%
\end{APACrefauthors}%
\unskip\
\newblock
\APACrefYearMonthDay{2018}{}{}.
\newblock
{\BBOQ}\APACrefatitle {Response of the Intertropical Convergence Zone to Climate Change: {L}ocation, Width, and Strength} {Response of the intertropical convergence zone to climate change: {L}ocation, width, and strength}.{\BBCQ}
\newblock
\APACjournalVolNumPages{Current Climate Change Reports}{4}{}{355--370}.
\PrintBackRefs{\CurrentBib}

\bibitem [\protect \citeauthoryear {%
Clough%
, Iacono%
\BCBL {}\ \BBA {} Moncet%
}{%
Clough%
\ \protect \BOthers {.}}{%
{\protect \APACyear {1992}}%
}]{%
clough1992line}
\APACinsertmetastar {%
clough1992line}%
\begin{APACrefauthors}%
Clough, S\BPBI A.%
, Iacono, M\BPBI J.%
\BCBL {}\ \BBA {} Moncet, J\BHBI L.%
\end{APACrefauthors}%
\unskip\
\newblock
\APACrefYearMonthDay{1992}{}{}.
\newblock
{\BBOQ}\APACrefatitle {Line-by-line calculations of atmospheric fluxes and cooling rates: Application to water vapor} {Line-by-line calculations of atmospheric fluxes and cooling rates: Application to water vapor}.{\BBCQ}
\newblock
\APACjournalVolNumPages{Journal of Geophysical Research: Atmospheres}{97}{D14}{15761--15785}.
\PrintBackRefs{\CurrentBib}

\bibitem [\protect \citeauthoryear {%
Cohen%
\ \BBA {} Pincus%
}{%
Cohen%
\ \BBA {} Pincus%
}{%
{\protect \APACyear {2025}}%
}]{%
cohen2025spectroscopic}
\APACinsertmetastar {%
cohen2025spectroscopic}%
\begin{APACrefauthors}%
Cohen, S.%
\BCBT {}\ \BBA {} Pincus, R.%
\end{APACrefauthors}%
\unskip\
\newblock
\APACrefYearMonthDay{2025}{}{}.
\newblock
{\BBOQ}\APACrefatitle {A spectroscopic theory for how mean rainfall changes with surface temperature} {A spectroscopic theory for how mean rainfall changes with surface temperature}.{\BBCQ}
\newblock
\APACjournalVolNumPages{Science Advances}{11}{19}{eadv6191}.
\PrintBackRefs{\CurrentBib}

\bibitem [\protect \citeauthoryear {%
Crisp%
, Fels%
\BCBL {}\ \BBA {} Schwarzkopf%
}{%
Crisp%
\ \protect \BOthers {.}}{%
{\protect \APACyear {1986}}%
}]{%
crisp1986approximate}
\APACinsertmetastar {%
crisp1986approximate}%
\begin{APACrefauthors}%
Crisp, D.%
, Fels, S\BPBI B.%
\BCBL {}\ \BBA {} Schwarzkopf, M.%
\end{APACrefauthors}%
\unskip\
\newblock
\APACrefYearMonthDay{1986}{}{}.
\newblock
{\BBOQ}\APACrefatitle {Approximate methods for finding CO2 15-$\mu$m band transmission in planetary atmospheres} {Approximate methods for finding co2 15-$\mu$m band transmission in planetary atmospheres}.{\BBCQ}
\newblock
\APACjournalVolNumPages{Journal of Geophysical Research: Atmospheres}{91}{D11}{11851--11866}.
\PrintBackRefs{\CurrentBib}

\bibitem [\protect \citeauthoryear {%
Davis%
\ \BBA {} Birner%
}{%
Davis%
\ \BBA {} Birner%
}{%
{\protect \APACyear {2022}}%
}]{%
davis2022eddy}
\APACinsertmetastar {%
davis2022eddy}%
\begin{APACrefauthors}%
Davis, N\BPBI A.%
\BCBT {}\ \BBA {} Birner, T.%
\end{APACrefauthors}%
\unskip\
\newblock
\APACrefYearMonthDay{2022}{}{}.
\newblock
{\BBOQ}\APACrefatitle {Eddy-mediated Hadley cell expansion due to axisymmetric angular momentum adjustment to greenhouse gas forcings} {Eddy-mediated hadley cell expansion due to axisymmetric angular momentum adjustment to greenhouse gas forcings}.{\BBCQ}
\newblock
\APACjournalVolNumPages{Journal of the Atmospheric Sciences}{79}{1}{141--159}.
\PrintBackRefs{\CurrentBib}

\bibitem [\protect \citeauthoryear {%
Dwyer%
\ \BBA {} O'Gorman%
}{%
Dwyer%
\ \BBA {} O'Gorman%
}{%
{\protect \APACyear {2017}}%
}]{%
dwyer17}
\APACinsertmetastar {%
dwyer17}%
\begin{APACrefauthors}%
Dwyer, J\BPBI G.%
\BCBT {}\ \BBA {} O'Gorman, P\BPBI A.%
\end{APACrefauthors}%
\unskip\
\newblock
\APACrefYearMonthDay{2017}{}{}.
\newblock
{\BBOQ}\APACrefatitle {Moist formulations of the {E}liassen--{P}alm flux and their connection to the surface westerlies} {Moist formulations of the {E}liassen--{P}alm flux and their connection to the surface westerlies}.{\BBCQ}
\newblock
\APACjournalVolNumPages{J. Atmos. Sci.}{74}{2}{513--530}.
\PrintBackRefs{\CurrentBib}

\bibitem [\protect \citeauthoryear {%
Feng%
, Paynter%
\BCBL {}\ \BBA {} Menzel%
}{%
Feng%
\ \protect \BOthers {.}}{%
{\protect \APACyear {2023}}%
}]{%
feng2023stable}
\APACinsertmetastar {%
feng2023stable}%
\begin{APACrefauthors}%
Feng, J.%
, Paynter, D.%
\BCBL {}\ \BBA {} Menzel, R.%
\end{APACrefauthors}%
\unskip\
\newblock
\APACrefYearMonthDay{2023}{}{}.
\newblock
{\BBOQ}\APACrefatitle {How a stable greenhouse effect on Earth is maintained under global warming} {How a stable greenhouse effect on earth is maintained under global warming}.{\BBCQ}
\newblock
\APACjournalVolNumPages{Journal of Geophysical Research: Atmospheres}{128}{9}{e2022JD038124}.
\PrintBackRefs{\CurrentBib}

\bibitem [\protect \citeauthoryear {%
Fildier%
, Muller%
, Pincus%
\BCBL {}\ \BBA {} Fueglistaler%
}{%
Fildier%
\ \protect \BOthers {.}}{%
{\protect \APACyear {2023}}%
}]{%
fildier2023moisture}
\APACinsertmetastar {%
fildier2023moisture}%
\begin{APACrefauthors}%
Fildier, B.%
, Muller, C.%
, Pincus, R.%
\BCBL {}\ \BBA {} Fueglistaler, S.%
\end{APACrefauthors}%
\unskip\
\newblock
\APACrefYearMonthDay{2023}{}{}.
\newblock
{\BBOQ}\APACrefatitle {How moisture shapes low-level radiative cooling in subsidence regimes} {How moisture shapes low-level radiative cooling in subsidence regimes}.{\BBCQ}
\newblock
\APACjournalVolNumPages{AGU Advances}{4}{3}{e2023AV000880}.
\PrintBackRefs{\CurrentBib}

\bibitem [\protect \citeauthoryear {%
Frierson%
}{%
Frierson%
}{%
{\protect \APACyear {2007}}%
}]{%
frierson07b}
\APACinsertmetastar {%
frierson07b}%
\begin{APACrefauthors}%
Frierson, D\BPBI M\BPBI W.%
\end{APACrefauthors}%
\unskip\
\newblock
\APACrefYearMonthDay{2007}{}{}.
\newblock
{\BBOQ}\APACrefatitle {The dynamics of idealized convection schemes and their effect on the zonally averaged tropical circulation} {The dynamics of idealized convection schemes and their effect on the zonally averaged tropical circulation}.{\BBCQ}
\newblock
\APACjournalVolNumPages{J. Atmos. Sci.}{64}{}{1959--1976}.
\PrintBackRefs{\CurrentBib}

\bibitem [\protect \citeauthoryear {%
Frierson%
, Held%
\BCBL {}\ \BBA {} Zurita-Gotor%
}{%
Frierson%
\ \protect \BOthers {.}}{%
{\protect \APACyear {2006}}%
}]{%
Frierson06a}
\APACinsertmetastar {%
Frierson06a}%
\begin{APACrefauthors}%
Frierson, D\BPBI M\BPBI W.%
, Held, I\BPBI M.%
\BCBL {}\ \BBA {} Zurita-Gotor, P.%
\end{APACrefauthors}%
\unskip\
\newblock
\APACrefYearMonthDay{2006}{}{}.
\newblock
{\BBOQ}\APACrefatitle {A gray-radiation aquaplanet moist {GCM}. {Part~I}: Static stability and eddy scale} {A gray-radiation aquaplanet moist {GCM}. {Part~I}: Static stability and eddy scale}.{\BBCQ}
\newblock
\APACjournalVolNumPages{J. Atmos. Sci.}{63}{}{2548--2566}.
\PrintBackRefs{\CurrentBib}

\bibitem [\protect \citeauthoryear {%
Geen%
, Czaja%
\BCBL {}\ \BBA {} Haigh%
}{%
Geen%
\ \protect \BOthers {.}}{%
{\protect \APACyear {2016}}%
}]{%
geen2016effects}
\APACinsertmetastar {%
geen2016effects}%
\begin{APACrefauthors}%
Geen, R.%
, Czaja, A.%
\BCBL {}\ \BBA {} Haigh, J\BPBI D.%
\end{APACrefauthors}%
\unskip\
\newblock
\APACrefYearMonthDay{2016}{}{}.
\newblock
{\BBOQ}\APACrefatitle {The effects of increasing humidity on heat transport by extratropical waves} {The effects of increasing humidity on heat transport by extratropical waves}.{\BBCQ}
\newblock
\APACjournalVolNumPages{Geophysical Research Letters}{43}{15}{8314--8321}.
\PrintBackRefs{\CurrentBib}

\bibitem [\protect \citeauthoryear {%
Goldblatt%
, Robinson%
, Zahnle%
\BCBL {}\ \BBA {} Crisp%
}{%
Goldblatt%
\ \protect \BOthers {.}}{%
{\protect \APACyear {2013}}%
}]{%
goldblatt2013low}
\APACinsertmetastar {%
goldblatt2013low}%
\begin{APACrefauthors}%
Goldblatt, C.%
, Robinson, T\BPBI D.%
, Zahnle, K\BPBI J.%
\BCBL {}\ \BBA {} Crisp, D.%
\end{APACrefauthors}%
\unskip\
\newblock
\APACrefYearMonthDay{2013}{}{}.
\newblock
{\BBOQ}\APACrefatitle {Low simulated radiation limit for runaway greenhouse climates} {Low simulated radiation limit for runaway greenhouse climates}.{\BBCQ}
\newblock
\APACjournalVolNumPages{Nature Geoscience}{6}{8}{661--667}.
\PrintBackRefs{\CurrentBib}

\bibitem [\protect \citeauthoryear {%
Gordon%
\ \protect \BOthers {.}}{%
Gordon%
\ \protect \BOthers {.}}{%
{\protect \APACyear {2017}}%
}]{%
gordon2017hitran2016}
\APACinsertmetastar {%
gordon2017hitran2016}%
\begin{APACrefauthors}%
Gordon, I\BPBI E.%
, Rothman, L\BPBI S.%
, Hill, C.%
, Kochanov, R\BPBI V.%
, Tan, Y.%
, Bernath, P\BPBI F.%
\BDBL {}others%
\end{APACrefauthors}%
\unskip\
\newblock
\APACrefYearMonthDay{2017}{}{}.
\newblock
{\BBOQ}\APACrefatitle {The HITRAN2016 molecular spectroscopic database} {The hitran2016 molecular spectroscopic database}.{\BBCQ}
\newblock
\APACjournalVolNumPages{Journal of quantitative spectroscopy and radiative transfer}{203}{}{3--69}.
\PrintBackRefs{\CurrentBib}

\bibitem [\protect \citeauthoryear {%
Guendelman%
\ \BBA {} Kaspi%
}{%
Guendelman%
\ \BBA {} Kaspi%
}{%
{\protect \APACyear {2020}}%
}]{%
guendelman20}
\APACinsertmetastar {%
guendelman20}%
\begin{APACrefauthors}%
Guendelman, I.%
\BCBT {}\ \BBA {} Kaspi, Y.%
\end{APACrefauthors}%
\unskip\
\newblock
\APACrefYearMonthDay{2020}{}{}.
\newblock
{\BBOQ}\APACrefatitle {Atmospheric dynamics on terrestrial planets with eccentric orbits} {Atmospheric dynamics on terrestrial planets with eccentric orbits}.{\BBCQ}
\newblock
\APACjournalVolNumPages{Astrophys. J.}{901}{}{46}.
\PrintBackRefs{\CurrentBib}

\bibitem [\protect \citeauthoryear {%
Held%
}{%
Held%
}{%
{\protect \APACyear {2005}}%
}]{%
held05b}
\APACinsertmetastar {%
held05b}%
\begin{APACrefauthors}%
Held, I\BPBI M.%
\end{APACrefauthors}%
\unskip\
\newblock
\APACrefYearMonthDay{2005}{}{}.
\newblock
{\BBOQ}\APACrefatitle {The gap between simulation and understanding in climate modeling} {The gap between simulation and understanding in climate modeling}.{\BBCQ}
\newblock
\APACjournalVolNumPages{Bull. Amer. Meteor. Soc.}{86}{}{1609--1614}.
\PrintBackRefs{\CurrentBib}

\bibitem [\protect \citeauthoryear {%
Hogan%
\ \BBA {} Matricardi%
}{%
Hogan%
\ \BBA {} Matricardi%
}{%
{\protect \APACyear {2020}}%
}]{%
hogan2020evaluating}
\APACinsertmetastar {%
hogan2020evaluating}%
\begin{APACrefauthors}%
Hogan, R\BPBI J.%
\BCBT {}\ \BBA {} Matricardi, M.%
\end{APACrefauthors}%
\unskip\
\newblock
\APACrefYearMonthDay{2020}{}{}.
\newblock
{\BBOQ}\APACrefatitle {Evaluating and improving the treatment of gases in radiation schemes: the Correlated K-Distribution Model Intercomparison Project (CKDMIP)} {Evaluating and improving the treatment of gases in radiation schemes: the correlated k-distribution model intercomparison project (ckdmip)}.{\BBCQ}
\newblock
\APACjournalVolNumPages{Geoscientific Model Development}{13}{12}{6501--6521}.
\PrintBackRefs{\CurrentBib}

\bibitem [\protect \citeauthoryear {%
Iacono%
\ \protect \BOthers {.}}{%
Iacono%
\ \protect \BOthers {.}}{%
{\protect \APACyear {2008}}%
}]{%
iacono2008radiative}
\APACinsertmetastar {%
iacono2008radiative}%
\begin{APACrefauthors}%
Iacono, M\BPBI J.%
, Delamere, J\BPBI S.%
, Mlawer, E\BPBI J.%
, Shephard, M\BPBI W.%
, Clough, S\BPBI A.%
\BCBL {}\ \BBA {} Collins, W\BPBI D.%
\end{APACrefauthors}%
\unskip\
\newblock
\APACrefYearMonthDay{2008}{}{}.
\newblock
{\BBOQ}\APACrefatitle {Radiative forcing by long-lived greenhouse gases: Calculations with the AER radiative transfer models} {Radiative forcing by long-lived greenhouse gases: Calculations with the aer radiative transfer models}.{\BBCQ}
\newblock
\APACjournalVolNumPages{Journal of Geophysical Research: Atmospheres}{113}{D13}{}.
\PrintBackRefs{\CurrentBib}

\bibitem [\protect \citeauthoryear {%
Jeevanjee%
\ \BBA {} Fueglistaler%
}{%
Jeevanjee%
\ \BBA {} Fueglistaler%
}{%
{\protect \APACyear {2020}}%
{\protect \APACexlab {{\protect \BCnt {1}}}}}]{%
jeevanjee2020cooling}
\APACinsertmetastar {%
jeevanjee2020cooling}%
\begin{APACrefauthors}%
Jeevanjee, N.%
\BCBT {}\ \BBA {} Fueglistaler, S.%
\end{APACrefauthors}%
\unskip\
\newblock
\APACrefYearMonthDay{2020{\protect \BCnt {1}}}{}{}.
\newblock
{\BBOQ}\APACrefatitle {On the cooling-to-space approximation} {On the cooling-to-space approximation}.{\BBCQ}
\newblock
\APACjournalVolNumPages{Journal of the Atmospheric Sciences}{77}{2}{465--478}.
\PrintBackRefs{\CurrentBib}

\bibitem [\protect \citeauthoryear {%
Jeevanjee%
\ \BBA {} Fueglistaler%
}{%
Jeevanjee%
\ \BBA {} Fueglistaler%
}{%
{\protect \APACyear {2020}}%
{\protect \APACexlab {{\protect \BCnt {2}}}}}]{%
jeevanjee2020simple}
\APACinsertmetastar {%
jeevanjee2020simple}%
\begin{APACrefauthors}%
Jeevanjee, N.%
\BCBT {}\ \BBA {} Fueglistaler, S.%
\end{APACrefauthors}%
\unskip\
\newblock
\APACrefYearMonthDay{2020{\protect \BCnt {2}}}{}{}.
\newblock
{\BBOQ}\APACrefatitle {Simple spectral models for atmospheric radiative cooling} {Simple spectral models for atmospheric radiative cooling}.{\BBCQ}
\newblock
\APACjournalVolNumPages{Journal of the Atmospheric Sciences}{77}{2}{479--497}.
\PrintBackRefs{\CurrentBib}

\bibitem [\protect \citeauthoryear {%
Jeevanjee%
, Hassanzadeh%
, Hill%
\BCBL {}\ \BBA {} Sheshadri%
}{%
Jeevanjee%
\ \protect \BOthers {.}}{%
{\protect \APACyear {2017}}%
}]{%
jeevanjee17}
\APACinsertmetastar {%
jeevanjee17}%
\begin{APACrefauthors}%
Jeevanjee, N.%
, Hassanzadeh, P.%
, Hill, S.%
\BCBL {}\ \BBA {} Sheshadri, A.%
\end{APACrefauthors}%
\unskip\
\newblock
\APACrefYearMonthDay{2017}{}{}.
\newblock
{\BBOQ}\APACrefatitle {A perspective on climate model hierarchies} {A perspective on climate model hierarchies}.{\BBCQ}
\newblock
\APACjournalVolNumPages{J. Adv. Model. Earth Syst.}{}{}{1760--1771}.
\PrintBackRefs{\CurrentBib}

\bibitem [\protect \citeauthoryear {%
Jeevanjee%
, Koll%
\BCBL {}\ \BBA {} Lutsko%
}{%
Jeevanjee%
, Koll%
\BCBL {}\ \BBA {} Lutsko%
}{%
{\protect \APACyear {2021}}%
}]{%
jeevanjee2021simpson}
\APACinsertmetastar {%
jeevanjee2021simpson}%
\begin{APACrefauthors}%
Jeevanjee, N.%
, Koll, D\BPBI D.%
\BCBL {}\ \BBA {} Lutsko, N.%
\end{APACrefauthors}%
\unskip\
\newblock
\APACrefYearMonthDay{2021}{}{}.
\newblock
{\BBOQ}\APACrefatitle {“Simpson's Law” and the spectral cancellation of climate feedbacks} {“simpson's law” and the spectral cancellation of climate feedbacks}.{\BBCQ}
\newblock
\APACjournalVolNumPages{Geophysical Research Letters}{48}{14}{e2021GL093699}.
\PrintBackRefs{\CurrentBib}

\bibitem [\protect \citeauthoryear {%
Jeevanjee%
, Seeley%
, Paynter%
\BCBL {}\ \BBA {} Fueglistaler%
}{%
Jeevanjee%
, Seeley%
\BCBL {}\ \protect \BOthers {.}}{%
{\protect \APACyear {2021}}%
}]{%
jeevanjee2021analytical}
\APACinsertmetastar {%
jeevanjee2021analytical}%
\begin{APACrefauthors}%
Jeevanjee, N.%
, Seeley, J\BPBI T.%
, Paynter, D.%
\BCBL {}\ \BBA {} Fueglistaler, S.%
\end{APACrefauthors}%
\unskip\
\newblock
\APACrefYearMonthDay{2021}{}{}.
\newblock
{\BBOQ}\APACrefatitle {An analytical model for spatially varying clear-sky CO 2 forcing} {An analytical model for spatially varying clear-sky co 2 forcing}.{\BBCQ}
\newblock
\APACjournalVolNumPages{Journal of Climate}{34}{23}{9463--9480}.
\PrintBackRefs{\CurrentBib}

\bibitem [\protect \citeauthoryear {%
S\BPBI M.~Kang%
, Held%
, Frierson%
\BCBL {}\ \BBA {} Zhao%
}{%
S\BPBI M.~Kang%
\ \protect \BOthers {.}}{%
{\protect \APACyear {2008}}%
}]{%
Kang08}
\APACinsertmetastar {%
Kang08}%
\begin{APACrefauthors}%
Kang, S\BPBI M.%
, Held, I\BPBI M.%
, Frierson, D\BPBI M\BPBI W.%
\BCBL {}\ \BBA {} Zhao, M.%
\end{APACrefauthors}%
\unskip\
\newblock
\APACrefYearMonthDay{2008}{}{}.
\newblock
{\BBOQ}\APACrefatitle {The response of the {ITCZ} to extratropical thermal forcing: Idealized slab-ocean experiments with a {GCM}} {The response of the {ITCZ} to extratropical thermal forcing: Idealized slab-ocean experiments with a {GCM}}.{\BBCQ}
\newblock
\APACjournalVolNumPages{J. Climate}{21}{}{3521--3532}.
\PrintBackRefs{\CurrentBib}

\bibitem [\protect \citeauthoryear {%
S\BPBI M.~Kang%
\ \BBA {} Polvani%
}{%
S\BPBI M.~Kang%
\ \BBA {} Polvani%
}{%
{\protect \APACyear {2011}}%
}]{%
kang2011interannual}
\APACinsertmetastar {%
kang2011interannual}%
\begin{APACrefauthors}%
Kang, S\BPBI M.%
\BCBT {}\ \BBA {} Polvani, L\BPBI M.%
\end{APACrefauthors}%
\unskip\
\newblock
\APACrefYearMonthDay{2011}{}{}.
\newblock
{\BBOQ}\APACrefatitle {The interannual relationship between the latitude of the eddy-driven jet and the edge of the Hadley cell} {The interannual relationship between the latitude of the eddy-driven jet and the edge of the hadley cell}.{\BBCQ}
\newblock
\APACjournalVolNumPages{Journal of Climate}{24}{2}{563--568}.
\PrintBackRefs{\CurrentBib}

\bibitem [\protect \citeauthoryear {%
W.~Kang%
\ \BBA {} Wordsworth%
}{%
W.~Kang%
\ \BBA {} Wordsworth%
}{%
{\protect \APACyear {2019}}%
}]{%
kang2019collapse}
\APACinsertmetastar {%
kang2019collapse}%
\begin{APACrefauthors}%
Kang, W.%
\BCBT {}\ \BBA {} Wordsworth, R.%
\end{APACrefauthors}%
\unskip\
\newblock
\APACrefYearMonthDay{2019}{}{}.
\newblock
{\BBOQ}\APACrefatitle {Collapse of the general circulation in shortwave-absorbing atmospheres: an idealized model study} {Collapse of the general circulation in shortwave-absorbing atmospheres: an idealized model study}.{\BBCQ}
\newblock
\APACjournalVolNumPages{The Astrophysical Journal Letters}{885}{1}{L18}.
\PrintBackRefs{\CurrentBib}

\bibitem [\protect \citeauthoryear {%
Kaspi%
\ \BBA {} Showman%
}{%
Kaspi%
\ \BBA {} Showman%
}{%
{\protect \APACyear {2015}}%
}]{%
kaspi15}
\APACinsertmetastar {%
kaspi15}%
\begin{APACrefauthors}%
Kaspi, Y.%
\BCBT {}\ \BBA {} Showman, A\BPBI P.%
\end{APACrefauthors}%
\unskip\
\newblock
\APACrefYearMonthDay{2015}{}{}.
\newblock
{\BBOQ}\APACrefatitle {Atmospheric dynamics of terrestrial exoplanets over a wide range of orbital and atmospheric parameters} {Atmospheric dynamics of terrestrial exoplanets over a wide range of orbital and atmospheric parameters}.{\BBCQ}
\newblock
\APACjournalVolNumPages{Astrophys. J.}{804}{}{60}.
\PrintBackRefs{\CurrentBib}

\bibitem [\protect \citeauthoryear {%
Kempton%
, Zahnle%
\BCBL {}\ \BBA {} Fortney%
}{%
Kempton%
\ \protect \BOthers {.}}{%
{\protect \APACyear {2011}}%
}]{%
kempton2011atmospheric}
\APACinsertmetastar {%
kempton2011atmospheric}%
\begin{APACrefauthors}%
Kempton, E\BPBI M\BHBI R.%
, Zahnle, K.%
\BCBL {}\ \BBA {} Fortney, J\BPBI J.%
\end{APACrefauthors}%
\unskip\
\newblock
\APACrefYearMonthDay{2011}{}{}.
\newblock
{\BBOQ}\APACrefatitle {The atmospheric chemistry of GJ 1214b: Photochemistry and clouds} {The atmospheric chemistry of gj 1214b: Photochemistry and clouds}.{\BBCQ}
\newblock
\APACjournalVolNumPages{The Astrophysical Journal}{745}{1}{3}.
\PrintBackRefs{\CurrentBib}

\bibitem [\protect \citeauthoryear {%
Kluft%
, Dacie%
, Brath%
, Buehler%
\BCBL {}\ \BBA {} Stevens%
}{%
Kluft%
\ \protect \BOthers {.}}{%
{\protect \APACyear {2021}}%
}]{%
kluft2021temperature}
\APACinsertmetastar {%
kluft2021temperature}%
\begin{APACrefauthors}%
Kluft, L.%
, Dacie, S.%
, Brath, M.%
, Buehler, S\BPBI A.%
\BCBL {}\ \BBA {} Stevens, B.%
\end{APACrefauthors}%
\unskip\
\newblock
\APACrefYearMonthDay{2021}{}{}.
\newblock
{\BBOQ}\APACrefatitle {Temperature-dependence of the clear-sky feedback in radiative-convective equilibrium} {Temperature-dependence of the clear-sky feedback in radiative-convective equilibrium}.{\BBCQ}
\newblock
\APACjournalVolNumPages{Geophysical Research Letters}{48}{22}{e2021GL094649}.
\PrintBackRefs{\CurrentBib}

\bibitem [\protect \citeauthoryear {%
Koll%
\ \BBA {} Cronin%
}{%
Koll%
\ \BBA {} Cronin%
}{%
{\protect \APACyear {2018}}%
}]{%
koll18}
\APACinsertmetastar {%
koll18}%
\begin{APACrefauthors}%
Koll, D\BPBI D\BPBI B.%
\BCBT {}\ \BBA {} Cronin, T\BPBI W.%
\end{APACrefauthors}%
\unskip\
\newblock
\APACrefYearMonthDay{2018}{}{}.
\newblock
{\BBOQ}\APACrefatitle {Earth’s outgoing longwave radiation linear due to H$_2$O greenhouse effect} {Earth’s outgoing longwave radiation linear due to h$_2$o greenhouse effect}.{\BBCQ}
\newblock
\APACjournalVolNumPages{Proc. Nat. Acad. Sci.}{115}{}{10293--10298}.
\PrintBackRefs{\CurrentBib}

\bibitem [\protect \citeauthoryear {%
Koll%
, Jeevanjee%
\BCBL {}\ \BBA {} Lutsko%
}{%
Koll%
\ \protect \BOthers {.}}{%
{\protect \APACyear {2023}}%
}]{%
koll23}
\APACinsertmetastar {%
koll23}%
\begin{APACrefauthors}%
Koll, D\BPBI D\BPBI B.%
, Jeevanjee, N.%
\BCBL {}\ \BBA {} Lutsko, N\BPBI J.%
\end{APACrefauthors}%
\unskip\
\newblock
\APACrefYearMonthDay{2023}{}{}.
\newblock
{\BBOQ}\APACrefatitle {An analytic model for the clear-sky longwave feedback} {An analytic model for the clear-sky longwave feedback}.{\BBCQ}
\newblock
\APACjournalVolNumPages{J. Atmos. Sci.}{80}{}{1923--1951}.
\PrintBackRefs{\CurrentBib}

\bibitem [\protect \citeauthoryear {%
Lachmy%
\ \BBA {} Shaw%
}{%
Lachmy%
\ \BBA {} Shaw%
}{%
{\protect \APACyear {2018}}%
}]{%
lachmy2018connecting}
\APACinsertmetastar {%
lachmy2018connecting}%
\begin{APACrefauthors}%
Lachmy, O.%
\BCBT {}\ \BBA {} Shaw, T.%
\end{APACrefauthors}%
\unskip\
\newblock
\APACrefYearMonthDay{2018}{}{}.
\newblock
{\BBOQ}\APACrefatitle {Connecting the energy and momentum flux response to climate change using the Eliassen--Palm relation} {Connecting the energy and momentum flux response to climate change using the eliassen--palm relation}.{\BBCQ}
\newblock
\APACjournalVolNumPages{Journal of Climate}{31}{18}{7401--7416}.
\PrintBackRefs{\CurrentBib}

\bibitem [\protect \citeauthoryear {%
Lewis%
\ \protect \BOthers {.}}{%
Lewis%
\ \protect \BOthers {.}}{%
{\protect \APACyear {2024}}%
}]{%
lewis2024assessing}
\APACinsertmetastar {%
lewis2024assessing}%
\begin{APACrefauthors}%
Lewis, N\BPBI T.%
, England, M\BPBI R.%
, Screen, J\BPBI A.%
, Geen, R.%
, Mudhar, R.%
, Seviour, W\BPBI J.%
\BCBL {}\ \BBA {} Thomson, S\BPBI I.%
\end{APACrefauthors}%
\unskip\
\newblock
\APACrefYearMonthDay{2024}{}{}.
\newblock
{\BBOQ}\APACrefatitle {Assessing the Spurious Impacts of Ice-Constraining Methods on the Climate Response to Sea Ice Loss Using an Idealized Aquaplanet GCM} {Assessing the spurious impacts of ice-constraining methods on the climate response to sea ice loss using an idealized aquaplanet gcm}.{\BBCQ}
\newblock
\APACjournalVolNumPages{Journal of Climate}{37}{24}{6729--6750}.
\PrintBackRefs{\CurrentBib}

\bibitem [\protect \citeauthoryear {%
Liu%
, Yang%
, Ding%
, Chen%
\BCBL {}\ \BBA {} Hu%
}{%
Liu%
\ \protect \BOthers {.}}{%
{\protect \APACyear {2024}}%
}]{%
liu2024hydrologic}
\APACinsertmetastar {%
liu2024hydrologic}%
\begin{APACrefauthors}%
Liu, J.%
, Yang, J.%
, Ding, F.%
, Chen, G.%
\BCBL {}\ \BBA {} Hu, Y.%
\end{APACrefauthors}%
\unskip\
\newblock
\APACrefYearMonthDay{2024}{}{}.
\newblock
{\BBOQ}\APACrefatitle {Hydrologic cycle weakening in hothouse climates} {Hydrologic cycle weakening in hothouse climates}.{\BBCQ}
\newblock
\APACjournalVolNumPages{Science Advances}{10}{17}{eado2515}.
\PrintBackRefs{\CurrentBib}

\bibitem [\protect \citeauthoryear {%
Lutsko%
\ \BBA {} Popp%
}{%
Lutsko%
\ \BBA {} Popp%
}{%
{\protect \APACyear {2018}}%
}]{%
lutsko2018influence}
\APACinsertmetastar {%
lutsko2018influence}%
\begin{APACrefauthors}%
Lutsko, N\BPBI J.%
\BCBT {}\ \BBA {} Popp, M.%
\end{APACrefauthors}%
\unskip\
\newblock
\APACrefYearMonthDay{2018}{}{}.
\newblock
{\BBOQ}\APACrefatitle {The influence of meridional gradients in insolation and longwave optical depth on the climate of a gray radiation GCM} {The influence of meridional gradients in insolation and longwave optical depth on the climate of a gray radiation gcm}.{\BBCQ}
\newblock
\APACjournalVolNumPages{Journal of Climate}{31}{19}{7803--7822}.
\PrintBackRefs{\CurrentBib}

\bibitem [\protect \citeauthoryear {%
Manabe%
\ \BBA {} Wetherald%
}{%
Manabe%
\ \BBA {} Wetherald%
}{%
{\protect \APACyear {1967}}%
}]{%
manabe67}
\APACinsertmetastar {%
manabe67}%
\begin{APACrefauthors}%
Manabe, S.%
\BCBT {}\ \BBA {} Wetherald, R\BPBI T.%
\end{APACrefauthors}%
\unskip\
\newblock
\APACrefYearMonthDay{1967}{}{}.
\newblock
{\BBOQ}\APACrefatitle {Thermal Equilibrium of the Atmosphere with a Given Distribution of Relative Humidity} {Thermal equilibrium of the atmosphere with a given distribution of relative humidity}.{\BBCQ}
\newblock
\APACjournalVolNumPages{J. Atmos. Sci.}{24}{}{241--259}.
\PrintBackRefs{\CurrentBib}

\bibitem [\protect \citeauthoryear {%
Match%
\ \BBA {} Fueglistaler%
}{%
Match%
\ \BBA {} Fueglistaler%
}{%
{\protect \APACyear {2021}}%
}]{%
match2021large}
\APACinsertmetastar {%
match2021large}%
\begin{APACrefauthors}%
Match, A.%
\BCBT {}\ \BBA {} Fueglistaler, S.%
\end{APACrefauthors}%
\unskip\
\newblock
\APACrefYearMonthDay{2021}{}{}.
\newblock
{\BBOQ}\APACrefatitle {Large internal variability dominates over global warming signal in observed lower stratospheric QBO amplitude} {Large internal variability dominates over global warming signal in observed lower stratospheric qbo amplitude}.{\BBCQ}
\newblock
\APACjournalVolNumPages{Journal of Climate}{34}{24}{9823--9836}.
\PrintBackRefs{\CurrentBib}

\bibitem [\protect \citeauthoryear {%
McKim%
, Jeevanjee%
, Vallis%
\BCBL {}\ \BBA {} Lewis%
}{%
McKim%
\ \protect \BOthers {.}}{%
{\protect \APACyear {2025}}%
}]{%
mckim2025water}
\APACinsertmetastar {%
mckim2025water}%
\begin{APACrefauthors}%
McKim, B\BPBI A.%
, Jeevanjee, N.%
, Vallis, G\BPBI K.%
\BCBL {}\ \BBA {} Lewis, N\BPBI T.%
\end{APACrefauthors}%
\unskip\
\newblock
\APACrefYearMonthDay{2025}{}{}.
\newblock
{\BBOQ}\APACrefatitle {Water vapor spectroscopy and thermodynamics constrain Earth's tropopause temperature} {Water vapor spectroscopy and thermodynamics constrain earth's tropopause temperature}.{\BBCQ}
\newblock
\APACjournalVolNumPages{AGU Advances}{6}{2}{e2024AV001206}.
\PrintBackRefs{\CurrentBib}

\bibitem [\protect \citeauthoryear {%
Merlis%
\ \BBA {} Schneider%
}{%
Merlis%
\ \BBA {} Schneider%
}{%
{\protect \APACyear {2010}}%
}]{%
merlis10}
\APACinsertmetastar {%
merlis10}%
\begin{APACrefauthors}%
Merlis, T\BPBI M.%
\BCBT {}\ \BBA {} Schneider, T.%
\end{APACrefauthors}%
\unskip\
\newblock
\APACrefYearMonthDay{2010}{}{}.
\newblock
{\BBOQ}\APACrefatitle {Atmospheric dynamics of {E}arth-like tidally locked aquaplanets} {Atmospheric dynamics of {E}arth-like tidally locked aquaplanets}.{\BBCQ}
\newblock
\APACjournalVolNumPages{J. Adv. Model. Earth Syst.}{2}{}{Art. \#13, 17 pp.}
\PrintBackRefs{\CurrentBib}

\bibitem [\protect \citeauthoryear {%
Mitchell%
\ \BBA {} Lora%
}{%
Mitchell%
\ \BBA {} Lora%
}{%
{\protect \APACyear {2016}}%
}]{%
mitchell2016climate}
\APACinsertmetastar {%
mitchell2016climate}%
\begin{APACrefauthors}%
Mitchell, J\BPBI L.%
\BCBT {}\ \BBA {} Lora, J\BPBI M.%
\end{APACrefauthors}%
\unskip\
\newblock
\APACrefYearMonthDay{2016}{}{}.
\newblock
{\BBOQ}\APACrefatitle {The climate of Titan} {The climate of titan}.{\BBCQ}
\newblock
\APACjournalVolNumPages{Annual Review of Earth and Planetary Sciences}{44}{1}{353--380}.
\PrintBackRefs{\CurrentBib}

\bibitem [\protect \citeauthoryear {%
Mlawer%
\ \protect \BOthers {.}}{%
Mlawer%
\ \protect \BOthers {.}}{%
{\protect \APACyear {2012}}%
}]{%
mlawer2012development}
\APACinsertmetastar {%
mlawer2012development}%
\begin{APACrefauthors}%
Mlawer, E\BPBI J.%
, Payne, V\BPBI H.%
, Moncet, J\BHBI L.%
, Delamere, J\BPBI S.%
, Alvarado, M\BPBI J.%
\BCBL {}\ \BBA {} Tobin, D\BPBI C.%
\end{APACrefauthors}%
\unskip\
\newblock
\APACrefYearMonthDay{2012}{}{}.
\newblock
{\BBOQ}\APACrefatitle {Development and recent evaluation of the MT\_CKD model of continuum absorption} {Development and recent evaluation of the mt\_ckd model of continuum absorption}.{\BBCQ}
\newblock
\APACjournalVolNumPages{Philosophical Transactions of the Royal Society A: Mathematical, Physical and Engineering Sciences}{370}{1968}{2520--2556}.
\PrintBackRefs{\CurrentBib}

\bibitem [\protect \citeauthoryear {%
Mlawer%
, Taubman%
, Brown%
, Iacono%
\BCBL {}\ \BBA {} Clough%
}{%
Mlawer%
\ \protect \BOthers {.}}{%
{\protect \APACyear {1997}}%
}]{%
mlawer97}
\APACinsertmetastar {%
mlawer97}%
\begin{APACrefauthors}%
Mlawer, E\BPBI J.%
, Taubman, S\BPBI J.%
, Brown, P\BPBI D.%
, Iacono, M\BPBI J.%
\BCBL {}\ \BBA {} Clough, S\BPBI A.%
\end{APACrefauthors}%
\unskip\
\newblock
\APACrefYearMonthDay{1997}{}{}.
\newblock
{\BBOQ}\APACrefatitle {Radiative transfer for inhomogeneous atmospheres: {RRTM}, a validated correlated-k model for the longwave} {Radiative transfer for inhomogeneous atmospheres: {RRTM}, a validated correlated-k model for the longwave}.{\BBCQ}
\newblock
\APACjournalVolNumPages{J. Geophys. Res.}{102}{}{16663--16682}.
\PrintBackRefs{\CurrentBib}

\bibitem [\protect \citeauthoryear {%
Mlynczak%
\ \protect \BOthers {.}}{%
Mlynczak%
\ \protect \BOthers {.}}{%
{\protect \APACyear {2016}}%
}]{%
mlynczak2016spectroscopic}
\APACinsertmetastar {%
mlynczak2016spectroscopic}%
\begin{APACrefauthors}%
Mlynczak, M\BPBI G.%
, Daniels, T\BPBI S.%
, Kratz, D\BPBI P.%
, Feldman, D\BPBI R.%
, Collins, W\BPBI D.%
, Mlawer, E\BPBI J.%
\BDBL {}others%
\end{APACrefauthors}%
\unskip\
\newblock
\APACrefYearMonthDay{2016}{}{}.
\newblock
{\BBOQ}\APACrefatitle {The spectroscopic foundation of radiative forcing of climate by carbon dioxide} {The spectroscopic foundation of radiative forcing of climate by carbon dioxide}.{\BBCQ}
\newblock
\APACjournalVolNumPages{Geophysical research letters}{43}{10}{5318--5325}.
\PrintBackRefs{\CurrentBib}

\bibitem [\protect \citeauthoryear {%
O'Gorman%
\ \BBA {} Schneider%
}{%
O'Gorman%
\ \BBA {} Schneider%
}{%
{\protect \APACyear {2008}}%
}]{%
OGorman08b}
\APACinsertmetastar {%
OGorman08b}%
\begin{APACrefauthors}%
O'Gorman, P\BPBI A.%
\BCBT {}\ \BBA {} Schneider, T.%
\end{APACrefauthors}%
\unskip\
\newblock
\APACrefYearMonthDay{2008}{}{}.
\newblock
{\BBOQ}\APACrefatitle {The hydrological cycle over a wide range of climates simulated with an idealized {GCM}} {The hydrological cycle over a wide range of climates simulated with an idealized {GCM}}.{\BBCQ}
\newblock
\APACjournalVolNumPages{J. Climate}{21}{}{3815--3832}.
\PrintBackRefs{\CurrentBib}

\bibitem [\protect \citeauthoryear {%
Petty%
}{%
Petty%
}{%
{\protect \APACyear {2006}}%
}]{%
petty2006first}
\APACinsertmetastar {%
petty2006first}%
\begin{APACrefauthors}%
Petty, G\BPBI W.%
\end{APACrefauthors}%
\unskip\
\newblock
\APACrefYear{2006}.
\newblock
\APACrefbtitle {A first course in atmospheric radiation} {A first course in atmospheric radiation}.
\newblock
\APACaddressPublisher{}{Sundog publishing}.
\PrintBackRefs{\CurrentBib}

\bibitem [\protect \citeauthoryear {%
Pierrehumbert%
}{%
Pierrehumbert%
}{%
{\protect \APACyear {2010}}%
}]{%
Pierrehumbert10}
\APACinsertmetastar {%
Pierrehumbert10}%
\begin{APACrefauthors}%
Pierrehumbert, R\BPBI T.%
\end{APACrefauthors}%
\unskip\
\newblock
\APACrefYear{2010}.
\newblock
\APACrefbtitle {Principles of Planetary Climate} {Principles of planetary climate}.
\newblock
\APACaddressPublisher{}{Cambridge University Press}.
\PrintBackRefs{\CurrentBib}

\bibitem [\protect \citeauthoryear {%
Pincus%
, Mlawer%
\BCBL {}\ \BBA {} Delamere%
}{%
Pincus%
\ \protect \BOthers {.}}{%
{\protect \APACyear {2019}}%
}]{%
pincus2019balancing}
\APACinsertmetastar {%
pincus2019balancing}%
\begin{APACrefauthors}%
Pincus, R.%
, Mlawer, E\BPBI J.%
\BCBL {}\ \BBA {} Delamere, J\BPBI S.%
\end{APACrefauthors}%
\unskip\
\newblock
\APACrefYearMonthDay{2019}{}{}.
\newblock
{\BBOQ}\APACrefatitle {Balancing accuracy, efficiency, and flexibility in radiation calculations for dynamical models} {Balancing accuracy, efficiency, and flexibility in radiation calculations for dynamical models}.{\BBCQ}
\newblock
\APACjournalVolNumPages{Journal of Advances in Modeling Earth Systems}{11}{10}{3074--3089}.
\PrintBackRefs{\CurrentBib}

\bibitem [\protect \citeauthoryear {%
Pincus%
\ \protect \BOthers {.}}{%
Pincus%
\ \protect \BOthers {.}}{%
{\protect \APACyear {2015}}%
}]{%
pincus2015radiative}
\APACinsertmetastar {%
pincus2015radiative}%
\begin{APACrefauthors}%
Pincus, R.%
, Mlawer, E\BPBI J.%
, Oreopoulos, L.%
, Ackerman, A\BPBI S.%
, Baek, S.%
, Brath, M.%
\BDBL {}others%
\end{APACrefauthors}%
\unskip\
\newblock
\APACrefYearMonthDay{2015}{}{}.
\newblock
{\BBOQ}\APACrefatitle {Radiative flux and forcing parameterization error in aerosol-free clear skies} {Radiative flux and forcing parameterization error in aerosol-free clear skies}.{\BBCQ}
\newblock
\APACjournalVolNumPages{Geophysical Research Letters}{42}{13}{5485--5492}.
\PrintBackRefs{\CurrentBib}

\bibitem [\protect \citeauthoryear {%
Popp%
, Schmidt%
\BCBL {}\ \BBA {} Marotzke%
}{%
Popp%
\ \protect \BOthers {.}}{%
{\protect \APACyear {2016}}%
}]{%
popp2016transition}
\APACinsertmetastar {%
popp2016transition}%
\begin{APACrefauthors}%
Popp, M.%
, Schmidt, H.%
\BCBL {}\ \BBA {} Marotzke, J.%
\end{APACrefauthors}%
\unskip\
\newblock
\APACrefYearMonthDay{2016}{}{}.
\newblock
{\BBOQ}\APACrefatitle {Transition to a moist greenhouse with CO2 and solar forcing} {Transition to a moist greenhouse with co2 and solar forcing}.{\BBCQ}
\newblock
\APACjournalVolNumPages{Nature communications}{7}{1}{10627}.
\PrintBackRefs{\CurrentBib}

\bibitem [\protect \citeauthoryear {%
Reed%
\ \protect \BOthers {.}}{%
Reed%
\ \protect \BOthers {.}}{%
{\protect \APACyear {2025}}%
}]{%
reed2025idealized}
\APACinsertmetastar {%
reed2025idealized}%
\begin{APACrefauthors}%
Reed, K\BPBI A.%
, Medeiros, B.%
, Jablonowski, C.%
, Simpson, I\BPBI R.%
, Voigt, A.%
\BCBL {}\ \BBA {} Wing, A\BPBI A.%
\end{APACrefauthors}%
\unskip\
\newblock
\APACrefYearMonthDay{2025}{}{}.
\newblock
{\BBOQ}\APACrefatitle {Why idealized models are more important than ever in Earth system science} {Why idealized models are more important than ever in earth system science}.{\BBCQ}
\newblock
\APACjournalVolNumPages{AGU Advances}{6}{4}{e2025AV001716}.
\PrintBackRefs{\CurrentBib}

\bibitem [\protect \citeauthoryear {%
Roemer%
, Buehler%
\BCBL {}\ \BBA {} Menang%
}{%
Roemer%
\ \protect \BOthers {.}}{%
{\protect \APACyear {2025}}%
}]{%
roemer2025think}
\APACinsertmetastar {%
roemer2025think}%
\begin{APACrefauthors}%
Roemer, F\BPBI E.%
, Buehler, S\BPBI A.%
\BCBL {}\ \BBA {} Menang, K\BPBI P.%
\end{APACrefauthors}%
\unskip\
\newblock
\APACrefYearMonthDay{2025}{}{}.
\newblock
{\BBOQ}\APACrefatitle {How to think about the clear-sky shortwave water vapor feedback} {How to think about the clear-sky shortwave water vapor feedback}.{\BBCQ}
\newblock
\APACjournalVolNumPages{npj Climate and Atmospheric Science}{8}{1}{1--10}.
\PrintBackRefs{\CurrentBib}

\bibitem [\protect \citeauthoryear {%
Romps%
, Seeley%
\BCBL {}\ \BBA {} Edman%
}{%
Romps%
\ \protect \BOthers {.}}{%
{\protect \APACyear {2022}}%
}]{%
romps2022forcing}
\APACinsertmetastar {%
romps2022forcing}%
\begin{APACrefauthors}%
Romps, D\BPBI M.%
, Seeley, J\BPBI T.%
\BCBL {}\ \BBA {} Edman, J\BPBI P.%
\end{APACrefauthors}%
\unskip\
\newblock
\APACrefYearMonthDay{2022}{}{}.
\newblock
{\BBOQ}\APACrefatitle {Why the forcing from carbon dioxide scales as the logarithm of its concentration} {Why the forcing from carbon dioxide scales as the logarithm of its concentration}.{\BBCQ}
\newblock
\APACjournalVolNumPages{Journal of Climate}{35}{13}{4027--4047}.
\PrintBackRefs{\CurrentBib}

\bibitem [\protect \citeauthoryear {%
Santer%
\ \protect \BOthers {.}}{%
Santer%
\ \protect \BOthers {.}}{%
{\protect \APACyear {2003}}%
}]{%
santer2003behavior}
\APACinsertmetastar {%
santer2003behavior}%
\begin{APACrefauthors}%
Santer, B\BPBI D.%
, Sausen, R.%
, Wigley, T.%
, Boyle, J\BPBI S.%
, AchutaRao, K.%
, Doutriaux, C.%
\BDBL {}others%
\end{APACrefauthors}%
\unskip\
\newblock
\APACrefYearMonthDay{2003}{}{}.
\newblock
{\BBOQ}\APACrefatitle {Behavior of tropopause height and atmospheric temperature in models, reanalyses, and observations: Decadal changes} {Behavior of tropopause height and atmospheric temperature in models, reanalyses, and observations: Decadal changes}.{\BBCQ}
\newblock
\APACjournalVolNumPages{Journal of Geophysical Research: Atmospheres}{108}{D1}{ACL--1}.
\PrintBackRefs{\CurrentBib}

\bibitem [\protect \citeauthoryear {%
Schneider%
, O'Gorman%
\BCBL {}\ \BBA {} Levine%
}{%
Schneider%
\ \protect \BOthers {.}}{%
{\protect \APACyear {2010}}%
}]{%
Schneider10}
\APACinsertmetastar {%
Schneider10}%
\begin{APACrefauthors}%
Schneider, T.%
, O'Gorman, P\BPBI A.%
\BCBL {}\ \BBA {} Levine, X\BPBI J.%
\end{APACrefauthors}%
\unskip\
\newblock
\APACrefYearMonthDay{2010}{}{}.
\newblock
{\BBOQ}\APACrefatitle {Water Vapor and the Dynamics of Climate Changes} {Water vapor and the dynamics of climate changes}.{\BBCQ}
\newblock
\APACjournalVolNumPages{Rev. Geophys.}{48}{}{RG3001}.
\PrintBackRefs{\CurrentBib}

\bibitem [\protect \citeauthoryear {%
Seeley%
\ \BBA {} Jeevanjee%
}{%
Seeley%
\ \BBA {} Jeevanjee%
}{%
{\protect \APACyear {2021}}%
}]{%
seeley2021h2o}
\APACinsertmetastar {%
seeley2021h2o}%
\begin{APACrefauthors}%
Seeley, J\BPBI T.%
\BCBT {}\ \BBA {} Jeevanjee, N.%
\end{APACrefauthors}%
\unskip\
\newblock
\APACrefYearMonthDay{2021}{}{}.
\newblock
{\BBOQ}\APACrefatitle {H2O windows and CO2 radiator fins: A clear-sky explanation for the peak in equilibrium climate sensitivity} {H2o windows and co2 radiator fins: A clear-sky explanation for the peak in equilibrium climate sensitivity}.{\BBCQ}
\newblock
\APACjournalVolNumPages{Geophysical Research Letters}{48}{4}{e2020GL089609}.
\PrintBackRefs{\CurrentBib}

\bibitem [\protect \citeauthoryear {%
Shine%
, Ptashnik%
\BCBL {}\ \BBA {} R{\"a}del%
}{%
Shine%
\ \protect \BOthers {.}}{%
{\protect \APACyear {2012}}%
}]{%
shine2012water}
\APACinsertmetastar {%
shine2012water}%
\begin{APACrefauthors}%
Shine, K\BPBI P.%
, Ptashnik, I\BPBI V.%
\BCBL {}\ \BBA {} R{\"a}del, G.%
\end{APACrefauthors}%
\unskip\
\newblock
\APACrefYearMonthDay{2012}{}{}.
\newblock
{\BBOQ}\APACrefatitle {The water vapour continuum: brief history and recent developments} {The water vapour continuum: brief history and recent developments}.{\BBCQ}
\newblock
\APACjournalVolNumPages{Surveys in geophysics}{33}{3}{535--555}.
\PrintBackRefs{\CurrentBib}

\bibitem [\protect \citeauthoryear {%
Simpson%
}{%
Simpson%
}{%
{\protect \APACyear {1928}}%
}]{%
simpson1928olr}
\APACinsertmetastar {%
simpson1928olr}%
\begin{APACrefauthors}%
Simpson, G\BPBI C.%
\end{APACrefauthors}%
\unskip\
\newblock
\APACrefYearMonthDay{1928}{}{}.
\newblock
{\BBOQ}\APACrefatitle {Some studies in terrestrial radiation} {Some studies in terrestrial radiation}.{\BBCQ}
\newblock
\APACjournalVolNumPages{Memoirs of the Royal Meteorological Society}{2}{16}{69-95}.
\PrintBackRefs{\CurrentBib}

\bibitem [\protect \citeauthoryear {%
Spaulding-Astudillo%
\ \BBA {} Mitchell%
}{%
Spaulding-Astudillo%
\ \BBA {} Mitchell%
}{%
{\protect \APACyear {2025}}%
}]{%
spaulding2025clear}
\APACinsertmetastar {%
spaulding2025clear}%
\begin{APACrefauthors}%
Spaulding-Astudillo, F\BPBI E.%
\BCBT {}\ \BBA {} Mitchell, J\BPBI L.%
\end{APACrefauthors}%
\unskip\
\newblock
\APACrefYearMonthDay{2025}{}{}.
\newblock
{\BBOQ}\APACrefatitle {Clear-sky convergence, water vapor spectroscopy, and the origin of tropical congestus clouds} {Clear-sky convergence, water vapor spectroscopy, and the origin of tropical congestus clouds}.{\BBCQ}
\newblock
\APACjournalVolNumPages{AGU Advances}{6}{1}{e2024AV001300}.
\PrintBackRefs{\CurrentBib}

\bibitem [\protect \citeauthoryear {%
Stevens%
\ \BBA {} Kluft%
}{%
Stevens%
\ \BBA {} Kluft%
}{%
{\protect \APACyear {2023}}%
}]{%
stevens2023colorful}
\APACinsertmetastar {%
stevens2023colorful}%
\begin{APACrefauthors}%
Stevens, B.%
\BCBT {}\ \BBA {} Kluft, L.%
\end{APACrefauthors}%
\unskip\
\newblock
\APACrefYearMonthDay{2023}{}{}.
\newblock
{\BBOQ}\APACrefatitle {A colorful look at climate sensitivity} {A colorful look at climate sensitivity}.{\BBCQ}
\newblock
\APACjournalVolNumPages{Atmospheric Chemistry and Physics}{23}{23}{14673--14689}.
\PrintBackRefs{\CurrentBib}

\bibitem [\protect \citeauthoryear {%
Tan%
, Lachmy%
\BCBL {}\ \BBA {} Shaw%
}{%
Tan%
\ \protect \BOthers {.}}{%
{\protect \APACyear {2019}}%
}]{%
tan19}
\APACinsertmetastar {%
tan19}%
\begin{APACrefauthors}%
Tan, Z.%
, Lachmy, O.%
\BCBL {}\ \BBA {} Shaw, T\BPBI A.%
\end{APACrefauthors}%
\unskip\
\newblock
\APACrefYearMonthDay{2019}{}{}.
\newblock
{\BBOQ}\APACrefatitle {The sensitivity of the jet stream response to climate change to radiative assumptions} {The sensitivity of the jet stream response to climate change to radiative assumptions}.{\BBCQ}
\newblock
\APACjournalVolNumPages{J. Adv. Model. Earth Syst.}{11}{}{934--956}.
\PrintBackRefs{\CurrentBib}

\bibitem [\protect \citeauthoryear {%
Vallis%
\ \protect \BOthers {.}}{%
Vallis%
\ \protect \BOthers {.}}{%
{\protect \APACyear {2018}}%
}]{%
vallis18}
\APACinsertmetastar {%
vallis18}%
\begin{APACrefauthors}%
Vallis, G\BPBI K.%
, Colyer, G.%
, Geen, R.%
, Gerber, E.%
, Jucker, M.%
, Maher, P.%
\BDBL {}Thomson, S\BPBI I.%
\end{APACrefauthors}%
\unskip\
\newblock
\APACrefYearMonthDay{2018}{}{}.
\newblock
{\BBOQ}\APACrefatitle {Isca, {v1.0}: {A} framework for the global modelling of the atmospheres of Earth and other planets at varying levels of complexity} {Isca, {v1.0}: {A} framework for the global modelling of the atmospheres of earth and other planets at varying levels of complexity}.{\BBCQ}
\newblock
\APACjournalVolNumPages{Geosci. Model Dev.}{11}{}{843--859}.
\PrintBackRefs{\CurrentBib}

\bibitem [\protect \citeauthoryear {%
Vallis%
, Zurita-Gotor%
, Cairns%
\BCBL {}\ \BBA {} Kidston%
}{%
Vallis%
\ \protect \BOthers {.}}{%
{\protect \APACyear {2015}}%
}]{%
vallis15}
\APACinsertmetastar {%
vallis15}%
\begin{APACrefauthors}%
Vallis, G\BPBI K.%
, Zurita-Gotor, P.%
, Cairns, C.%
\BCBL {}\ \BBA {} Kidston, J.%
\end{APACrefauthors}%
\unskip\
\newblock
\APACrefYearMonthDay{2015}{}{}.
\newblock
{\BBOQ}\APACrefatitle {Response of the large-scale structure of the atmosphere to global warming} {Response of the large-scale structure of the atmosphere to global warming}.{\BBCQ}
\newblock
\APACjournalVolNumPages{Quart. J. Roy. Meteor. Soc.}{}{}{}.
\PrintBackRefs{\CurrentBib}

\bibitem [\protect \citeauthoryear {%
Virtanen%
\ \protect \BOthers {.}}{%
Virtanen%
\ \protect \BOthers {.}}{%
{\protect \APACyear {2020}}%
}]{%
virtanen2020scipy}
\APACinsertmetastar {%
virtanen2020scipy}%
\begin{APACrefauthors}%
Virtanen, P.%
, Gommers, R.%
, Oliphant, T\BPBI E.%
, Haberland, M.%
, Reddy, T.%
, Cournapeau, D.%
\BDBL {}others%
\end{APACrefauthors}%
\unskip\
\newblock
\APACrefYearMonthDay{2020}{}{}.
\newblock
{\BBOQ}\APACrefatitle {SciPy 1.0: fundamental algorithms for scientific computing in Python} {Scipy 1.0: fundamental algorithms for scientific computing in python}.{\BBCQ}
\newblock
\APACjournalVolNumPages{Nature methods}{17}{3}{261--272}.
\PrintBackRefs{\CurrentBib}

\bibitem [\protect \citeauthoryear {%
Waugh%
\ \protect \BOthers {.}}{%
Waugh%
\ \protect \BOthers {.}}{%
{\protect \APACyear {2018}}%
}]{%
waugh2018revisiting}
\APACinsertmetastar {%
waugh2018revisiting}%
\begin{APACrefauthors}%
Waugh, D\BPBI W.%
, Grise, K\BPBI M.%
, Seviour, W\BPBI J.%
, Davis, S\BPBI M.%
, Davis, N.%
, Adam, O.%
\BDBL {}others%
\end{APACrefauthors}%
\unskip\
\newblock
\APACrefYearMonthDay{2018}{}{}.
\newblock
{\BBOQ}\APACrefatitle {Revisiting the relationship among metrics of tropical expansion} {Revisiting the relationship among metrics of tropical expansion}.{\BBCQ}
\newblock
\APACjournalVolNumPages{Journal of Climate}{31}{18}{7565--7581}.
\PrintBackRefs{\CurrentBib}

\bibitem [\protect \citeauthoryear {%
Weaver%
\ \BBA {} Ramanathan%
}{%
Weaver%
\ \BBA {} Ramanathan%
}{%
{\protect \APACyear {1995}}%
}]{%
weaver1995deductions}
\APACinsertmetastar {%
weaver1995deductions}%
\begin{APACrefauthors}%
Weaver, C.%
\BCBT {}\ \BBA {} Ramanathan, V.%
\end{APACrefauthors}%
\unskip\
\newblock
\APACrefYearMonthDay{1995}{}{}.
\newblock
{\BBOQ}\APACrefatitle {Deductions from a simple climate model: Factors governing surface temperature and atmospheric thermal structure} {Deductions from a simple climate model: Factors governing surface temperature and atmospheric thermal structure}.{\BBCQ}
\newblock
\APACjournalVolNumPages{Journal of Geophysical Research: Atmospheres}{100}{D6}{11585--11591}.
\PrintBackRefs{\CurrentBib}

\bibitem [\protect \citeauthoryear {%
Wills%
, Levine%
\BCBL {}\ \BBA {} Schneider%
}{%
Wills%
\ \protect \BOthers {.}}{%
{\protect \APACyear {2017}}%
}]{%
wills17}
\APACinsertmetastar {%
wills17}%
\begin{APACrefauthors}%
Wills, R\BPBI C.%
, Levine, X\BPBI J.%
\BCBL {}\ \BBA {} Schneider, T.%
\end{APACrefauthors}%
\unskip\
\newblock
\APACrefYearMonthDay{2017}{}{}.
\newblock
{\BBOQ}\APACrefatitle {Local energetic constraints on {W}alker circulation strength} {Local energetic constraints on {W}alker circulation strength}.{\BBCQ}
\newblock
\APACjournalVolNumPages{J. Atmos. Sci.}{74}{}{1907--1922}.
\PrintBackRefs{\CurrentBib}

\bibitem [\protect \citeauthoryear {%
Wilson%
\ \BBA {} Gea-Banacloche%
}{%
Wilson%
\ \BBA {} Gea-Banacloche%
}{%
{\protect \APACyear {2012}}%
}]{%
wilson2012simple}
\APACinsertmetastar {%
wilson2012simple}%
\begin{APACrefauthors}%
Wilson, D\BPBI J.%
\BCBT {}\ \BBA {} Gea-Banacloche, J.%
\end{APACrefauthors}%
\unskip\
\newblock
\APACrefYearMonthDay{2012}{}{}.
\newblock
{\BBOQ}\APACrefatitle {Simple model to estimate the contribution of atmospheric CO2 to the Earth’s greenhouse effect} {Simple model to estimate the contribution of atmospheric co2 to the earth’s greenhouse effect}.{\BBCQ}
\newblock
\APACjournalVolNumPages{American Journal of Physics}{80}{4}{306--315}.
\PrintBackRefs{\CurrentBib}

\bibitem [\protect \citeauthoryear {%
Wordsworth%
, Seeley%
\BCBL {}\ \BBA {} Shine%
}{%
Wordsworth%
\ \protect \BOthers {.}}{%
{\protect \APACyear {2024}}%
}]{%
wordsworth2024fermi}
\APACinsertmetastar {%
wordsworth2024fermi}%
\begin{APACrefauthors}%
Wordsworth, R.%
, Seeley, J.%
\BCBL {}\ \BBA {} Shine, K.%
\end{APACrefauthors}%
\unskip\
\newblock
\APACrefYearMonthDay{2024}{}{}.
\newblock
{\BBOQ}\APACrefatitle {Fermi resonance and the quantum mechanical basis of global warming} {Fermi resonance and the quantum mechanical basis of global warming}.{\BBCQ}
\newblock
\APACjournalVolNumPages{The Planetary Science Journal}{5}{3}{67}.
\PrintBackRefs{\CurrentBib}

\end{thebibliography}

%Reference citation instructions and examples:
%
% Please use ONLY \cite and \citeA for reference citations.
% \cite for parenthetical references
% ...as shown in recent studies (Simpson et al., 2019)
% \citeA for in-text citations
% ...Simpson et al. (2019) have shown...
%
%
%...as shown by \citeA{jskilby}.
%...as shown by \citeA{lewin76}, \citeA{carson86}, \citeA{bartoldy02}, and \citeA{rinaldi03}.
%...has been shown \cite{jskilbye}.
%...has been shown \cite{lewin76,carson86,bartoldy02,rinaldi03}.
%... \cite <i.e.>[]{lewin76,carson86,bartoldy02,rinaldi03}.
%...has been shown by \cite <e.g.,>[and others]{lewin76}.
%
% apacite uses < > for prenotes and [ ] for postnotes
% DO NOT use other cite commands (e.g., \citet, \citep, \citeyear, \nocite, \citealp, etc.).
%

\end{document}